%% file: SglmmProj_arxiv.tex
\documentclass[12pt]{article}
\usepackage{epsf}
\usepackage{geometry}
\usepackage{setspace}
\usepackage{algorithm}
\usepackage{color}
\usepackage[normalem]{ulem}
\usepackage{slashbox}
\usepackage{amsmath} 
\usepackage{standalone}
\usepackage{authblk}
\usepackage{pdflscape}
\usepackage{graphicx}
\usepackage{subfig}
\usepackage[comma,authoryear]{natbib}
\newgeometry{margin=1in}
\title{A Computationally Efficient Projection-Based Approach for Spatial Generalized Linear Mixed Models}
\author[1]{Yawen Guan}
\author[2]{Murali Haran}
\affil[1]{The Statistical and Applied Mathematical Sciences Institute, and  \newline Department of Statistics, 
	North Carolina State University,
	Raleigh, NC 27695, U.S.A.
	Email: yguan8@ncsu.edu}
\affil[2]{Department of Statistics,
	Pennsylvania State University,
	University Park, PA 16803, U.S.A.
	Email: muh10@psu.edu}

\date{}
\newcommand{\bs}[1]{\boldsymbol{#1}}

\begin{document}
\maketitle
\begin{abstract}
Inference for spatial generalized linear mixed models (SGLMMs) for high-dimensional non-Gaussian spatial data is computationally intensive. The computational challenge is due to the high-dimensional random effects and because Markov chain Monte Carlo (MCMC) algorithms for these models tend to be slow mixing. Moreover, spatial confounding inflates the variance of fixed effect (regression coefficient) estimates. Our approach addresses both the computational and confounding issues by replacing the high-dimensional spatial random effects with a reduced-dimensional representation based on random projections. Standard MCMC algorithms mix well and the reduced-dimensional setting speeds up computations per iteration. We show, via simulated examples, that Bayesian inference for this reduced-dimensional approach works well both in terms of inference as well as prediction; our methods also compare favorably to existing “reduced-rank” approaches. We also apply our methods to two real world data examples, one on bird count data and the other classifying rock types. 
\end{abstract}

\include{SglmmProjMain_secondrevision_arxiv}

\textbf{Acknowledgments: }We are grateful to Professor Sanjay Srinivasan at the Department of Energy and Mineral Engineering at Penn State University for providing the seismic data set, and to John Hughes, Jim Hodges and Ephraim Hanks for helpful discussions. This work was partially supported by the National Science Foundation through (1) NSF-DMS-1418090, (2) Network for Sustainable Climate Risk Management (SCRiM) under NSF cooperative agreement GEO1240507 and (3) DMS-1638521 to the Statistical and Applied Mathematical Sciences Institute. MH was partially supported by (1), (2) and (3); YG was partially supported by (1) and (3). We would also like to acknowledge the high-performance computing support from Yellowstone (ark:/85065/d7wd3xhc) provided by NCAR's Computational and Information Systems Laboratory, sponsored by the National Science Foundation. 


\appendix
\include{SglmmProjSupp_arxiv}

\newpage
\begin{singlespace}
\bibliography{Reference}
\end{singlespace}
\end{document}

%% file: SglmmProjMain_secondrevision_arxiv.tex
\section{Introduction}
Gaussian and non-Gaussian spatial data arise in a number of disciplines, for example, species counts in ecology, tree presence-absence data, and disease incidence data. Models for such data are important for scientific applications, for instance when fitting spatial regression models or when interpolating observations across continuous spatial domains.  Spatial generalized linear mixed models (SGLMMs) are popular and flexible models for spatial  data, both for continuous spatial domain or ``point-referenced" data \citep{Diggle1998}, where the spatial dependence is captured by random effects modeled using a Gaussian process, as well as for lattice or areal data \citep[cf.][]{Besag1991,rue2005gaussian} where dependence is captured via random effects modeled with Gaussian Markov random fields. SGLMMs have become very popular in a wide range of disciplines. In practice, however, SGLMMs pose some computational and inferential challenges: (i) computational issues due to high-dimensional random effects that are typically strongly cross-correlated -- these often result in slow mixing Markov chain Monte Carlo (MCMC) algorithms; (ii) computations involving large matrices; and (iii) spatial confounding between fixed and random effects, which can lead to variance-inflated estimation of regression coefficients \citep{RHZ2006,HughesHaran,ephraim2015}. In this manuscript we provide an approach for reducing the dimensions of the spatial random effects in SGLMM models. Our approach simultaneously addresses both computational issues as well as the confounding issue.

There is a large literature on fast computational methods for spatial models \citep[cf.][among many others]{Cressie2008,Banerjee2008,higdon1998convolution, shaby2012tapered,datta2015hierarchical}.  These methods have been very useful in practice, but they largely focus on linear (Gaussian) spatial models and do not consider the spatial confounding issue.  The predictive process approach \citep{Banerjee2008} has been an important contribution to the literature, and has also been studied in the SGLMM context. However, the predictive process approach requires that users provide reference knots, which can be challenging to specify; our method is more automated.   We also find that in some cases we obtain similar performance to the predictive process at far lower computational cost. Crucially, our approach is also able to easily address the spatial confounding issue. INLA \citep{rue2009inla} provides a sophisticated numerical approximation approach for SGLMMs. As we later discuss, INLA may be used in combination with the projection-based reparameterization approach we develop in this manuscript. This is useful for addressing confounding while also reducing computational costs. 

Restricted spatial regression models for areal and point-referenced spatial data \citep{RHZ2006,ephraim2015} address the confounding issue. However, these models are computationally intensive for large data sets. For areal data, \citet{HughesHaran} alleviate confounding in a computationally efficient manner by proposing a reparameterization that utilizes the underlying graph to reduce the dimension of random effects. To our knowledge, no existing approach alleviates spatial confounding and is computationally efficient for point-referenced non-Gaussian data. In this manuscript we describe a novel method that utilizes the principal components of covariance matrices to achieve fast computation for fitting traditional SGLMMs as well as restricted spatial regression. 

Our method relies on the random projections algorithm \citep{A.Banerjee2012,Sarlos2006,halko2011}, which allows a fast approximation of the leading eigencomponents. We show how we can build upon this projection-based approach to address the computational and inferential challenges of SGLMMs. The outline of the remainder of the paper is as follows. In Section \ref{sec:spatial}, we introduce spatial linear mixed models and explain how a generalized linear model formulation of these models is appropriate for non-Gaussian observations. We also examine the computational challenges and some current approaches. In Section \ref{sec:confounding}, we explain spatial confounding, how it affects interpretation of regression parameters, and describe how to alleviate confounding via orthogonalization. In Section \ref{sec:RP}, we describe our projection-based approach for both the continuous domain and lattice case. We study the inference and prediction performance of the proposed method via a simulation study in Section \ref{sec:simu}, and study our method in the context of applications in Section \ref{sec:application}. We conclude with a discussion of our work in Section \ref{sec:discuss}.

\section{Spatial Models}\label{sec:spatial}
\subsection{Spatial linear models}

Let $ Y(\bs{s}) $ denote an observation, and $ \bs{x}(\bs{s}) $ a $p-$dimensional vector of covariates at location $ \bs{s} $ in a spatial domain $ \mathcal{D} \subseteq \mathcal{R}^d$, where $d$ is typically 2 or 3. Given data locations $ \mathcal{S} = \left\lbrace \bs{s}_1,\dots,\bs{s}_n\right\rbrace $, the observations $ \bs{Y} = [Y(\bs{s}_1),...,Y(\bs{s}_n)]^T $ may show residual spatial structure after controlling for $  {X} = [\bs{x}(\bs{s}_1) ,...,\bs{x}(\bs{s}_n)]^T $. This can be taken into account by including spatially dependent random effects $ W(\bs{s}) $ to model the residual dependence, 
 \begin{equation}\label{eqn:slm}
		\begin{aligned}
			Y(\bs{s})&= \bs{x}(\bs{s})^T\bs{\beta} + W(\bs{s}) + \epsilon(\bs{s}),
		\end{aligned}
\end{equation} 
where $ \bs{\beta} $ are regression parameters. $ \left\lbrace \epsilon(\bs{s}): \bs{s} \in \mathcal{D}\right\rbrace $ is a small-scale (nugget) spatial effect/measurement error process, modeled as an uncorrelated Gaussian process with mean 0 and variance $\tau^2$. For point-referenced data, the random effects $ \left\lbrace W(\bs{s}):\bs{s} \in \mathcal{D}\right\rbrace  $ are typically modeled by a zero-mean stationary Gaussian process with a positive definite covariance function $ C(\cdot)$. Hence, for a finite set of locations, $ \bs{W} = [W(\bs{s_1}),\dots,W(\bs{s_n})]^T $ follows a multivariate normal distribution $\text{MVN}(\bs{0},\Sigma)$, with $ \Sigma_{ij} = \text{cov}(W(\bs{s}_i),W(\bs{s}_j)) = C(\vert\vert\bs{s_i},\bs{s_j}\vert\vert)$.


A commonly used class of covariance functions, assuming stationarity and isotropy, is the Mat\'{e}rn class \citep{Stein1999},
 \begin{equation*}
		C(\bs{s}_i,\bs{s}_j) = C(h) = \sigma^2 \rho(h;\phi,\nu)= \sigma^2 \frac{1}{\Gamma(\nu)2^{\nu-1}} \left( 
		\sqrt{2\nu} \frac{h}{\phi}\right)^\nu K_{\nu}\left( 
		\sqrt{2\nu} \frac{h}{\phi}\right),
\end{equation*} 
where $h=||\bs{s}_i-\bs{s}_j||$ denotes the Euclidean distance between pairs of locations, $ \sigma^2 $ is a variance parameter, and $ \rho $ is a positive definite correlation function parameterized by $ \phi $, the spatial range parameter, and $ \nu $, the smoothness parameter. $ \Gamma(\cdot) $ is the gamma function, and $K_{\nu} (\cdot)$ is the modified Bessel function of the second kind.

\subsection{Spatial generalized linear mixed models}

A popular way to model spatial non-Gaussian data is by using spatial generalized linear mixed models (SGLMMs) \citep[cf.][]{Diggle1998,Muralibook}. Let $ \left\lbrace Z(\bs{s}): \bs{s}\in \mathcal{D}\right\rbrace $ denote a non-Gaussian spatial field, and $g\left\lbrace \cdot \right\rbrace$ a known link function. Then, the conditional mean, $ E[Z(\bs{s})\mid\bs{\beta},W(\bs{s})] $ may be modeled as
 \begin{equation}\label{eqn:eta}
		\eta(\bs{s})  \equiv  g\left\lbrace E[Z(\bs{s})\mid\bs{\beta},W(\bs{s})] \right\rbrace = \bs{x}(\bs{s})^T \bs{\beta} + W(\bs{s}), \hspace{4mm} \bs{s} \in \mathcal{D}.
\end{equation} 
Conditional on $W(\bs{s})$, $\bs{Z}=[Z(\bs{s}_1),\dots,Z(\bs{s}_n)]^T$ are mutually independent, following a classical generalized linear model \citep[cf.][]{Diggle1998}. We provide two commonly used examples of SGLMMs for spatial binary and count data to illustrate our projection-based approach, the Poisson with log link and binary with logit link respectively. The projection-based approach presented in this paper generalizes to other link functions and observation models, as well as to cases where an additional nugget term is added to the model (\ref{eqn:eta}) \citep[cf.][]{Berrett2016}. Details for the nugget model are provided in the supplement S.5.

\subsection{Model fitting and computational challenges} 

The hierarchical structure of spatial models makes it convenient to use a Bayesian inferential approach. Often in practice, we fix the value of $ \nu $ and assign prior, $ p{(\bs{\theta,\beta})} $, to parameters $ \bs{\theta}$ and $\bs{\beta} $ where $ \bs{\theta} = (\phi,\sigma^2, \tau^2)^T $, then use  Markov Chain Monte Carlo (MCMC) to sample from the  posterior $\pi(\bs{\beta},\bs{\theta},\bs{W}\mid\bs{Z}) $. Fitting SGLMMs generally requires the evaluation of an n-dimension multivariate normal likelihood for every MCMC iteration, with matrix operations of order $n^3$ floating point operations (flops).
There are often strong correlations between the fixed and random effects \citep{Hodges2010}, and strong cross-correlations among the spatially dependent random effects. It is well known that this dependence is often an important cause of poor mixing in standard MCMC algorithms \citep[cf.][]{RobustMCMC2006,Haran2003,rue2005gaussian}. Furthermore, when the data locations are near each other, the covariance matrix may be near singular, resulting in numerical  instabilities \citep{A.Banerjee2012}.  These issues motivate the development of our reduced-dimensional approach to inference for SGLMMs. 

Considerable work has been done to address the above issues in the linear case,
where model inference and prediction are based on the marginal distribution $ \bs{Y}\mid\bs{\beta},\phi,\sigma^2,\tau^2 \sim \text{MVN}(X\bs{\beta}, \Sigma + \tau^2I)$. Several methods rely on low rank approximations or multi-resolution approaches to reduce computations involving the $ n\times n$ covariance matrix $ \Sigma $ \citep[cf.][]{A.Banerjee2012,sang2012,multiresolution,Cressie2008}. However, these methods do not readily extend to SGLMMs because they assume that the random effects may be ``marginalized out" in closed form. For SGLMMs, this is generally not possible. \citet{Sengupta2013,Sengupta2016} extended low rank approximation method to SGLMM setting. They used bi-square basis functions to represent random effects with addition random noise to capture fine-scale-variation. Their model for the spatial random effects has the form $ W(\bs{s}) = \bs{B}(\bs{s})^T\bs{\delta} + \epsilon(\bs{s}) $, where $ \bs{B}(\cdot) $ denotes the basis functions, $ \bs{\delta}\sim N(0,\Sigma_{\delta}) $ is a vector of random effects with unknown $ \Sigma_{\delta} $, and independent Gaussian noise $ \epsilon(\cdot) $. Due to the high dimension of the random effects, the authors proposed empirical-Bayesian inference, which combines Laplace approximations in an expectation-maximization (EM) algorithm to obtain parameter estimates. A notable exception is the predictive process approach \citep{Banerjee2008}, where the extension to SGLMMs has been well studied. This approach replaces random effect $ \bs{W} $ by  $\bs{W}^\ast $, the realization of $ W(\bs{s}) $ at $m$ $(<< n)$ reference knots $ S^\ast = \left\lbrace \bs{s}_1^\ast,\dots,\bs{s}_m^\ast \right\rbrace$; $ \bs{W}\approx C(\bs{s},\bs{s}^\ast)C^{\ast-1}\bs{W}^\ast$, where $ C^{\ast} =C(\bs{s}^\ast,\bs{s}^\ast) $ denotes the $ m\times m $ covariance matrix $\text{var}(\bs{W}^\ast)$. Correspondingly, $\Sigma$ is approximated by a low rank matrix $ c^T C^{\ast-1}c $, where $ c^T = C(\bs{s},\bs{s}^\ast) $ denotes the covariance matrix $ \text{cov}(\bs{W},\bs{W}^\ast) $. This method can be applied to both the linear and the generalized case. However, \cite{finley2009} points out that the predictive process underestimates the variance of $ \bs{W} $ and proposed a modified predictive process by defining $ W_{mod}(\bs{s}) = C(\bs{s},\bs{s}^\ast)C^{\ast-1}\bs{W}^\ast + \tilde{\epsilon}(\bs{s})$, where $ \tilde{\epsilon}(\bs{s}) \stackrel{ind.}{\sim} \text{N}(0,C(\bs{s},\bs{s})-\bs{c}^T(\bs{s},\bs{s}^\ast)C^{\ast-1}\bs{c}(\bs{s},\bs{s}^\ast) ) $. For the linear case, this adjustment adds little extra computation. However, for the SGLMM case, the modified predictive process puts us back to working with a high-dimensional random effect $\bs{W}_{mod}$. Furthermore, determining the number and placement of reference knots is a non-trivial challenge \citep[see][for some potential strategies]{finley2009}.

Another challenge with SGLMMs arises from the strong-correlations among random effects, which often results in poor Markov chain mixing. Reparameterization techniques \citep{RobustMCMC2006} can help with mixing; however, for high-dimensional spatial data, the reparameterizing step is computationally expensive and may not result in fast mixing. 

\section{Confounding and Restricted Spatial Regression}\label{sec:confounding}

Spatial confounding occurs when the spatially observed covariates are collinear with the spatial random effects. This is a common problem for both point-referenced and areal data \citep[cf.][]{ephraim2015, RHZ2006}. Here we demonstrate the confounding problem in a continuous spatial domain. Let $ \bs{\eta} = [\eta(\bs{s}_1),\dots,\eta(\bs{s}_n)]^T $ denote the transformed site-specific conditional means, where $ \eta(\bs{s}_i) = g\left\lbrace E[Z(\bs{s}_i)\mid\bs{\beta},W(\bs{s}_i)]\right\rbrace  $. The SGLMM is then 
 \begin{equation}\label{eqn:sglmm}
		\bs{\eta} = X\bs{\beta} + \bs{W}, \hspace{4mm}\bs{W}\sim MVN(0,\sigma^2R(\phi)),
\end{equation} 
where the covariance $\Sigma$ is  $\sigma^2R(\phi)$, with $R(\phi)$ a positive definite correlation matrix,   $R_{ij}(\phi) = \rho(||\bs{s}_i-\bs{s}_j||;\phi)$. $X$ are spatially observed covariates that may explain the random field of interest. $ \bs{W}$ is used as a smoothing device. When both $ X $ and $ \bs{W} $ are spatially smooth, they are often collinear \citep[cf.][]{ephraim2015}. This confounding problem may lead to variance inflation of the fixed effects \citep{Hodges2010}. 

Let $ P_{[X]} $ and $ P^{\perp}_{[X]} $ denote orthogonal projections onto the space spanned by $ X $ and its complement, respectively. Model (\ref{eqn:sglmm}) can be equivalently written as
 
	\begin{equation}\label{eqn:sglmmrsr}
		\bs{\eta}=X \bs{\beta} + P_{[X]}\bs{W} +P^{\perp}_{[X]}\bs{W}= X\left[\bs{\beta} + (X^TX)^{-1}X^T\bs{W}\right] + P^{\perp}_{[X]}\bs{W}
		= X\tilde{\bs{\beta}} + P^{\perp}_{[X]}\bs{W}.
\end{equation} In some cases, it may be reasonable to fit model (\ref{eqn:sglmmrsr}) to address the confounding issue by restricting the random effects to be orthogonal to the fixed effects in $X$ \citep{RHZ2006,HughesHaran,ephraim2015}. We refer to this as restricted spatial regression (RSR) in the remaining sections. After fitting the RSR via MCMC, we can obtain valid inference for $ \bs{\beta} $ using an \textit{a posteriori} adjustment based on the MCMC samples \citep{ephraim2015}. Let $k$ indicate  the $ k^{th} $ MCMC sample, then 
 \begin{equation}\label{eqn:adjust}
		\bs{\beta}^{(k)} = \bs{\tilde{\beta}}^{(k)} - (X^TX)^{-1}X^T \bs{W}^{(k)}.
\end{equation} 


Fitting RSR is just as computationally expensive as regular SGLMMs because the dimension of random effects $P_{[X]}^\perp\bs{W} $ remains large; their strong correlations lead to slow MCMC mixing. 

\section{Reducing Dimensions through Projection}\label{sec:RP}

Instead of working with the original size of the random effects $ \bs{W} $ in the model, we consider a reduced dimensional approximation. We want to reduce the dimension of random effects from $n$ to $m$ so that: (i) for a fixed $m$, the approximation to the original process comes close to minimizing the variance of the truncation error (details below),  (ii) the $m$ random effects
are nearly uncorrelated, and (iii) we reduce the number of random effects as far
as possible in order to reduce the dimensionality of the posterior distribution.

Let $\bs{\delta}$ denote a vector of the reduced-dimensional reparameterized random effects. The main idea of our approach is to obtain $\bs{\delta}$ by projecting $\bs{W}$ to its first-$m$ principal direction $V_m = [\bs{v_1},\dots,\bs{v_m}]$, and scaling by its eigenvalues $\Lambda_m = \text{diag}(\lambda_1,\dots,\lambda_m)$. Conditional on $\phi$, $\bs{W}$ is multivariate normal. Hence, $ \bs{\delta}=( V_m\Lambda_m^{-1/2})^T\bs{W}$ is conditionally independent given $\phi$,  ($\bs{\delta}\mid \phi \sim N(\bs{0},I)$). This reparameterization utilizing principal components minimizes the variance of the truncation error (details are provided below), decorrelates the random effects and reduces its dimension to $m$. However, exact eigendecomposition is computationally infeasible for high-dimensional observations, so we approximate the eigencomponents, $U_m\approx V_m$ and $D_m\approx\Lambda_m$, using a recently developed stochastic matrix approximation. Because the eigencomponents are approximated reasonably well by the random projections algorithm as illustrated in Section \ref{sec:rpapprox}, our reparameterization, $\bs{\delta} =( U_mD_m^{-1/2})^T\bs{W}$, is therefore close to the one based on the exact eigendecomposition. In the remainder of this section we describe the motivation and properties of the reparameterization approach. We describe random projections for fast approximations of the eigencomponents and illustrate its approximation performance. We then explain how to fit the reparameterized SGLMMs with random projection. We conclude this section by showing how our approach is also applicable to areal data, and compare it to the method in \citet{HughesHaran}.

Our reparameterized random effects model achieves (i)-(iii) above. Consider the spatial process $\left\lbrace W(\bs{s}): \bs{s} \in D \right\rbrace$ in (\ref{eqn:slm}) defined on a compact subset $D$ of $\mathcal{R}^d$. Let $ \left\lbrace \psi_i(\bs{s}): i = 1, \dots, \infty\right\rbrace$ and $ \left\lbrace \lambda_i: i = 1, \dots, \infty\right\rbrace$ be orthonormal eigenfunctions and eigenvalues, respectively, of the covariance function $C(\cdot)$ of $W(\bs{s})$. By Mercer's theorem, they satisfy $C(\bs{s},\bs{s}') = \sum_{i=1}^{\infty} \lambda_i \psi_i(\bs{s})\psi_i(\bs{s}')$ \citep{Adler1990}. By the Karhunen-Lo\`{e}ve (K-L) expansion, we can write $W(\bs{s}) = \sum_{i=1}^{\infty} \xi_i\sqrt{\lambda_i} \psi_i(\bs{s}),$ where $\left\lbrace \xi_i: i = 1, \dots, \infty\right\rbrace $ are orthonormal Gaussian \citep{Adler1990}. Assuming the eigenvalues are in descending order, $\lambda_1 \ge \lambda_2 \ge...,$ the truncated K-L expansion for $ W(\bs{s}) $, $\widetilde{W}(\bs{s}) = \sum_{i=1}^{m} \xi_i\sqrt{\lambda_i} \psi_i(\bs{s})$, minimizes the mean square error, $||W-\widetilde{W}||$, among all basis sets of order $ m $ \citep[][]{A.Banerjee2012,cressie2015statistics}. A discrete analogue for the truncated expansion of the process realization $\bs{W}$ is similar to the above. $\bs{W}$ has expansion $\bs{W}=\sum_{i=1}^{n} \xi_i\sqrt{\lambda_i}\bs{v}_i$, while its rank-$m$ approximation is $\bs{\widetilde{W}}=\sum_{i=1}^{m}\xi_i\sqrt{\lambda_i}\bs{v}_i$, where  $\left\lbrace\lambda_i,\bs{v}_i\right\rbrace$ are eigen-pairs of $\Sigma$. Let $V_n=[\bs{v}_1,\dots,\bs{v}_n]$ denote an $n\times n$ matrix of eigenvectors, and $\Lambda_n=\text{diag}(\lambda_1,\dots,\lambda_n)$ an $n\times n$ diagonal matrix of eigenvalues. We define $V_m= [\bs{v}_1,\dots,\bs{v}_m]$ and $\Lambda_m=\text{diag}(\lambda_1,\dots,\lambda_m)$ similarly. Then, the truncated expansion has variance $\text{var}(\widetilde{\bs{W}})=\widetilde{\Sigma} =V_m\Lambda_mV_m^T$, and it minimizes the variance of the truncation error $||\Sigma - \widetilde{\Sigma}|| = \sum_{i=m+1}^{n} \lambda_i$ \citep{A.Banerjee2012}.

Our reparameterization is a principal component analysis (PCA) based approach with the advantage that a reasonably small rank $m$ captures most of the spatial variation. For instance, we later demonstrate in a simulated example of data size $n$=1,000, rank $m$=50 is sufficient to achieve reasonable performance. We discuss heuristics for choosing an appropriate value for $m$ in Section \ref{sec:rank}. 

\subsection{Random projection}
Random projection is an approach that facilitates fast approximations of matrix operations \citep[see][and references therein]{halko2011}. Here we use it to approximate the principal components of covariance matrices. Before introducing the random matrix approach, we first describe a deterministic approach to approximate eigendecomposition. Various algorithms that  approximate eigencomponents using a submatrix of the original matrix are compared in \cite{homrighausen2016nystrom}; In our implementation, we used the Nystr\"{o}m method \citep{Williams01,drineas2005nystrom}. Let $K$ denote an $n\times n$ positive definite matrix to be decomposed; we can further denote its partition as $ K = \left[ \begin{array}{cc}
K_{11} & K_{12}\\K_{21}& K_{22}
\end{array} \right] $, where $K_{11}$ is $k\times k$ dimensions. The central idea of Nystr\"{o}m's method is to compute exact eigendecomposition on the lower-dimensional submatrix $K_{11}$, then use the resulting lower-dimensional eigencomponents to approximate eigencomponents of $K$. Let $V_c(A)$ and $\Lambda_c(A)$ be matrices of the first $c$ eigenvectors and eigenvalues, respectively, of a positive definite matrix A; therefore, both have $c$ columns. Then $\Lambda_k(K)$ is approximated by $\widetilde{\Lambda}_k(K)=\frac{n}{k}\Lambda_k(K_{11})$, a $k\times k$ diagonal matrix whose elements are the approximated eigenvalues and $V_k(K)$ is approximated by scaling $V_k(K_{11})$ up to high dimensions via
 \begin{equation*}
		\widetilde{V}_k(K) = \sqrt{\frac{k}{n}}\left[ \begin{array}{c}
			K_{11} \\K_{21} 
		\end{array} \right] V_k(K_{11})\Lambda_k(K_{11})^{-1} = \sqrt{\frac{k}{n}} \left[ \begin{array}{c}
			V_k(K_{11}) \\K_{21}V_k(K_{11})\Lambda_k(K_{11})^{-1}
		\end{array} \right].
\end{equation*}  From the Nystr\"{o}m method, we also obtain an approximation to $K$ by $ \widetilde{K} = \widetilde{V}_k(K)\widetilde{\Lambda}_k(K)\widetilde{V}_k(K)^T = \left[ \begin{array}{c}
K_{11} \\K_{21} 
\end{array} \right] K_{11}^{-1} \left[ K_{11}, K_{12}\right] $. The error in the approximation is $ \mid\mid K_{22} - K_{21}K_{11}^{-1}K_{12} \mid\mid$, which reflects the information lost from truncating $ K $ \citep{belabbas2009spectral}. However, the approximated eigenvectors above are not guaranteed to be orthogonal, hence we adopt a slight variant of the form \citep[similar to Algorithm 5.5 in][]{halko2011}. Let $C$ denote $\left[ \begin{array}{c}
K_{11} \\K_{21} 
\end{array} \right] V_k(K_{11})\Lambda_k(K_{11})^{-1/2}$; then, its singular value decomposition is $\text{SVD}(C) = U(C)D(C)V(C)^T$, where $D(C)$ is a $k\times k$ diagonal matrix with elements equal to the non-zero singular values, $U(C)$ is an $n\times k$ matrix of the left singular vectors, and $V(C)$ is a $k\times k$ matrix of the right singular vectors of $C$.  (From here on, we suppress the dependencies of $U$ and $D$ on $C$.) Then $\widetilde{K}$ can also be expressed as $CC^T = UD^2U^T$, satisfying $\widetilde{K}U = UD^2$. Therefore, $U$ and $D^2$ are the first $k$ eigencomponents of $\widetilde{K}$, and they are used as our approximation to the eigencomponents of $K$, respectively. This Nystr\"{o}m's approximation is summarized as step 2 in Algorithm \ref{RPalgorithm}. 

Nystr\"{o}m's method obtains the column space of $K$ from subsampling its columns $ {K \Phi = \left[ \begin{array}{c}
	K_{11} \\K_{21} 
	\end{array} \right]}$, where $\Phi$ is an $n\times k$ matrix by permuting the rows of $[I_{k\times k}, 0_{k\times (n-k)}]^T$; once $\Phi$ is fixed, the approximation is deterministic. Alternatives to approximate the column space of $K$ involving randomness have been proposed, such as weighted random subsampling of the rows and columns of $K$ by \cite{frieze2004fast}, subsampled randomized Hadamard transform by \cite{tropp2011improved} or a random projection matrix with \textit{iid} elements \citep{halko2011,bingham2001random,A.Banerjee2012}. Here we adopt the latter method with \textit{iid} Gaussian random variables. 

Rather than truncating $K$, we take $\Phi$ to be $K^{\alpha}\Omega$, where $\Omega$ is a $n\times k$ random matrix with $\Omega_{ij} \sim N(0,1/\sqrt{k} )$, and $\alpha=0, 1, \text{or } 2$ takes a small non-negative integer value for improving approximation (see a comparison for $\alpha$ in Section \ref{sec:rpapprox}). Then $K\Phi$ is randomly weighted linear combination of columns of $K$, by construction, it approximates the column space of K. The random matrix $\Omega$ is a low-dimensional embedding: $R^{n\times n} \to R^{n\times k}$, that satisfies Johnson-Lindenstrauss's transformation; it has a low distortion such that $\mid\vert\vert\Omega^Tv\vert\vert-\vert\vert v\vert\vert\mid$ is small for all $v\in V\subset \mathcal{R}^n$ with high probability \citep[for details on the embedding, we refer the readers to][]{dasgupta2003elementary}. Taking small powers of $K$ in the projection matrix $\Phi=K^{\alpha}\Omega$, enhances our approximation performance but this involves a tradeoff in terms of computational speed; in our implementation, we see substantial improvement by taking $\alpha = 1 \text{ or } 2$. We let $k=m+l$, where $m$ denotes the target rank, and $l$ is an oversampling factor typically set to 5 or 10 to reduce approximation error \citep{halko2011}. In our implementation, we noticed that taking small $l$ is not enough to give a good approximation of the eigenvectors corresponding to the smaller eigenvalues; therefore, we take $l=m$. The random projection approach to approximate the column space of K is summarized as step 1 in Algorithm \ref{RPalgorithm}.

In the context of Gaussian process regression, \citet{A.Banerjee2012} used a similar random projection algorithm to approximate the covariance matrix $\Sigma$. Here we will directly approximate the eigencomponents of the correlation matrix $R(\phi)$, because multiplying $\sigma^2$ does not affect the approximation. In our MCMC implementation, we obtain $U_m$ and $D_m$, the approximated leading $m$ eigencomponents of $R(\phi)$, using Algorithm \ref{RPalgorithm} for every $\phi$ value. Then we can obtain reparameterized random effects $ \bs{\delta}\mid\phi=(U_mD_m^{-1/2})^T\bs{W}$.

\begin{algorithm}
	\caption{Random projection algorithm:} 
	\label{RPalgorithm}
	
	Given a positive semi-definite matrix K, this algorithm approximates the leading $ m $ eigencomponents of $ K $ by utilizing Nystr\"{o}m's method.
	\begin{enumerate}
		\item Low dimensional projection from $R^{n\times n}$ to $R^{n\times k}$, $ m < k << n $: \\
		Form $\Phi=K\Omega$ where $\Omega \in R^{n\times k}$ with $\Omega_{ij} = N(0,1/\sqrt{k})$
		\item Nystr\"{o}m's method to approximate eigendecomposition : \\
		Form $K_{11} = \Phi^T K\Phi$ \\
		SVD for $K_{11}$: $V_k(K_{11})\Lambda_k(K_{11})V_k(K_{11})^T$\\
		Form Nystr\"{o}m extension $C =  [K\Phi] [V_k(K_{11})\Lambda_k(K_{11})^{-1/2}]$\\
		SVD for $C$: $ UD V^T$
		\item Take the first $ m $ columns of $ U $, and the first $ m $ diagonal elements of $ D^2 $ as our approximation to the leading $ m $ eigencomponents of $ K $
	\end{enumerate}
\end{algorithm}

\subsection{Approximation comparison}\label{sec:rpapprox}
\begin{figure}[hpb]
	\centering
	\subfloat[$\nu$=0.5, $r$=0.5]{\includegraphics[width=6cm,height=5cm]{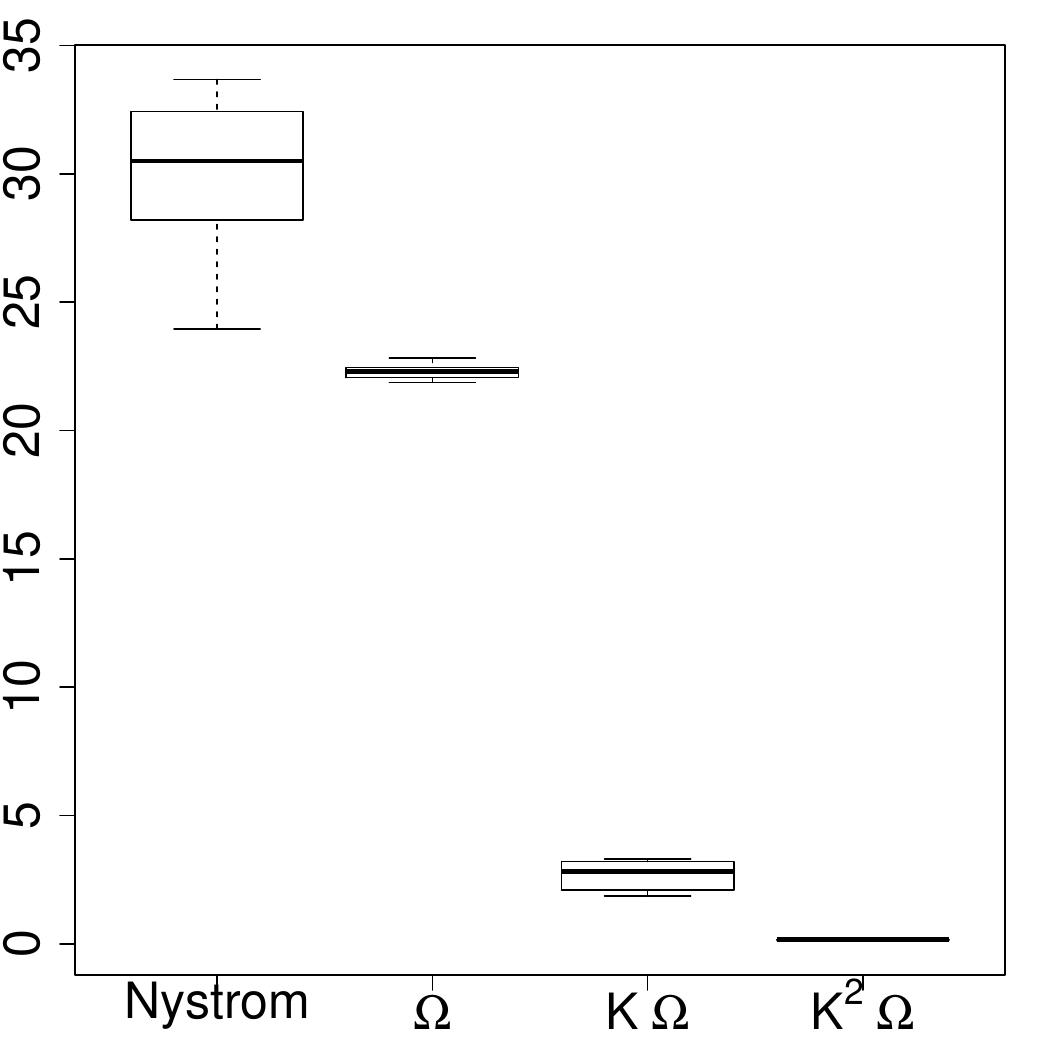}}
	\subfloat[$\nu$=0.5, $r$=0.5]{\includegraphics[width=6cm,height=5cm]{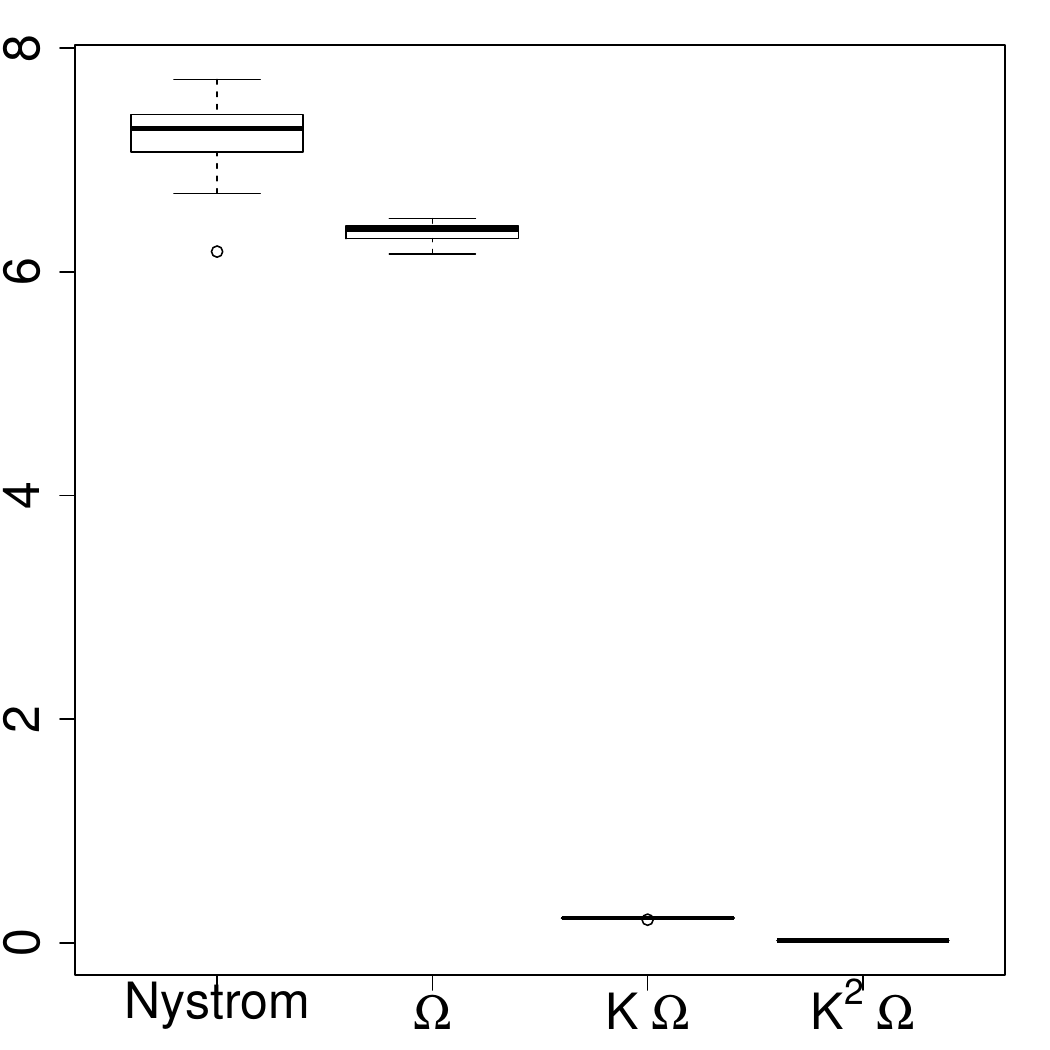}}
	
	\subfloat[$\nu$=2.5, $r$=0.5]{\includegraphics[width=6cm,height=5cm]{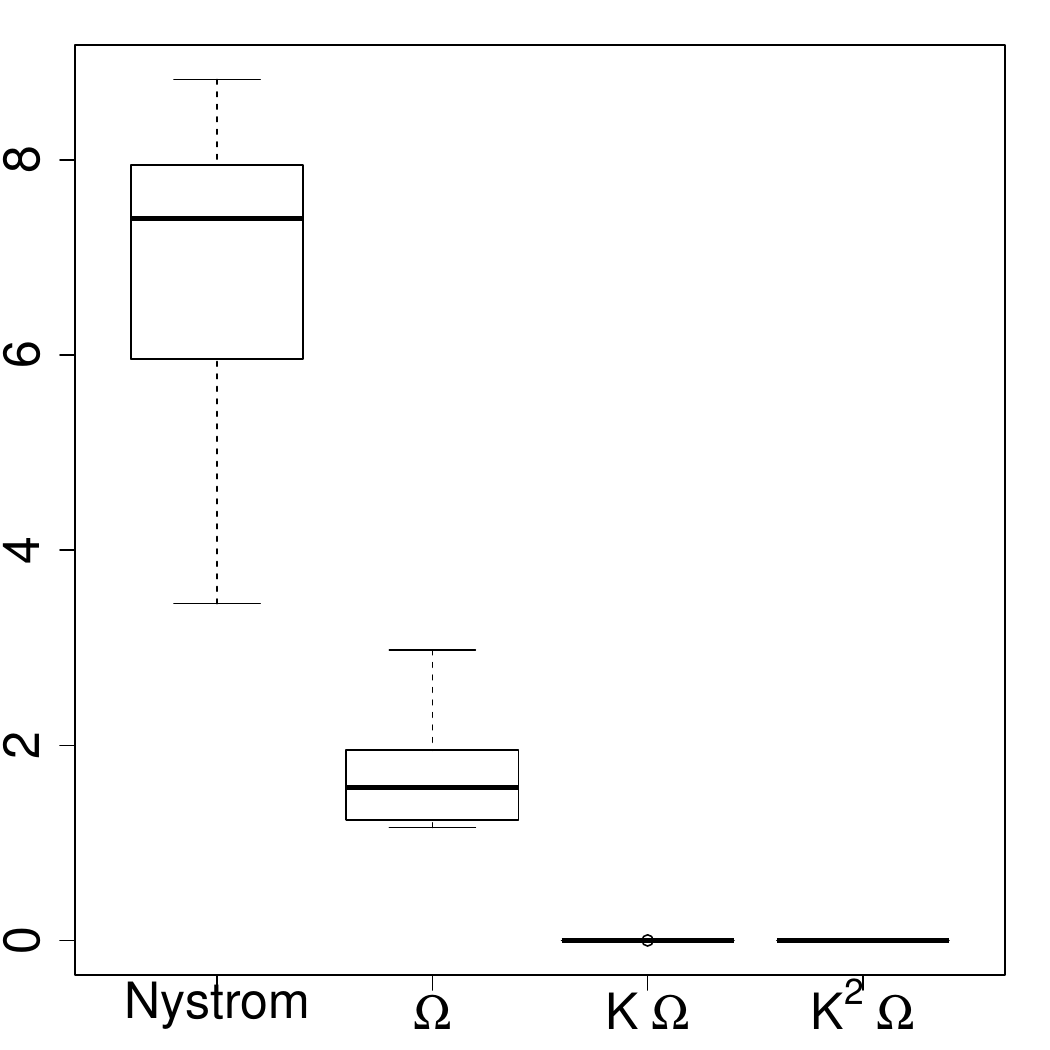}}
	\subfloat[$\nu$=2.5, $r$=0.5]{\includegraphics[width=6cm,height=5cm]{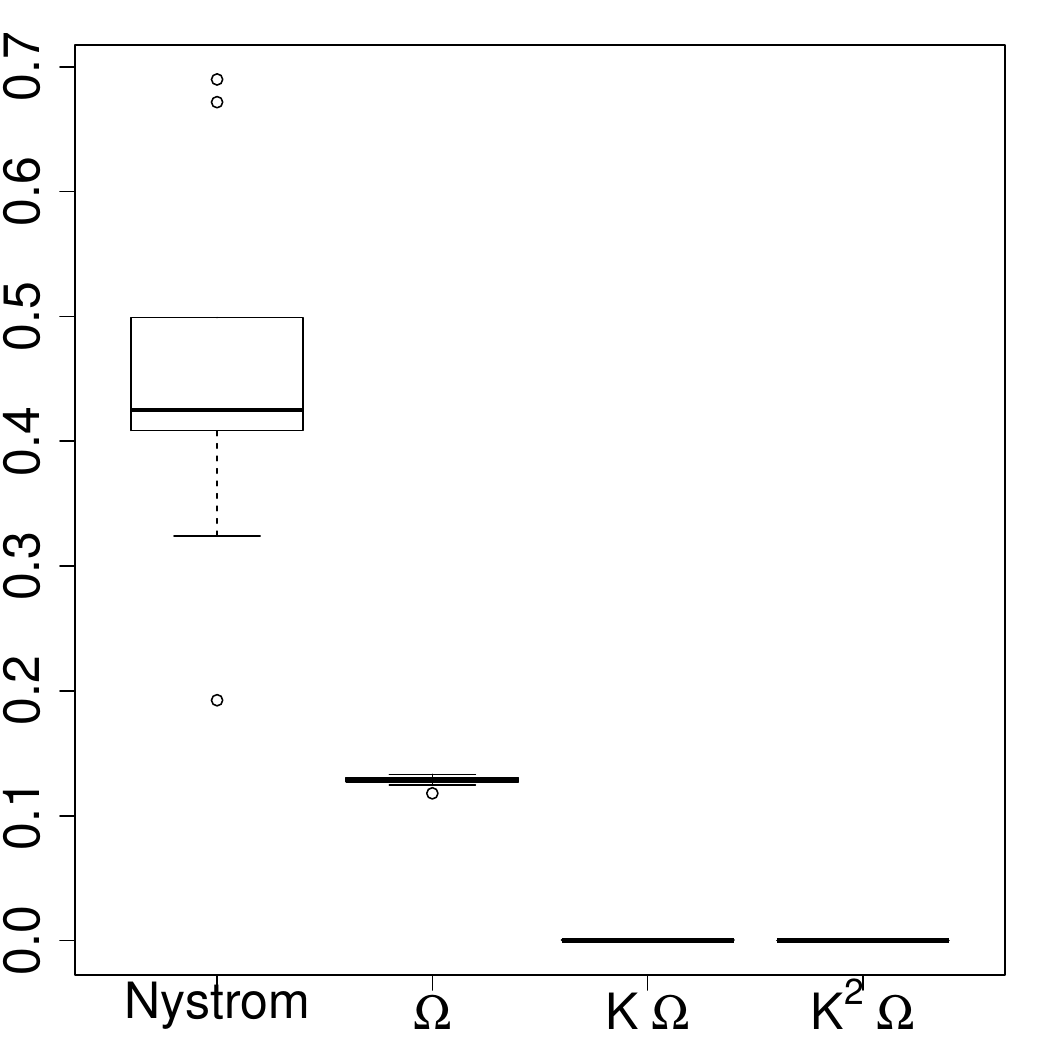}}	
	\caption{Approximation performance comparison shows that introducing randomness in $\Phi=K^\alpha\Omega$ improves the Nyst\"{o}m approximation to the eigenvectors (left column) and eigenvalues (right column). Letting $ \Phi $ to be $ K^\alpha\Omega $ with small power $ \alpha =1,2 $ also provides better approximation for both $\nu=0.5$ and $\nu=2.5$ with an effective range $ r=0.5 $.}
	\label{fig:aproxperformance}
\end{figure}

To illustrate the performance of introducing randomness in approximating eigencomponents using Nyst\"{o}m's method as described in Section 4.1. we perform a numerical experiment to compare the approximation performance of the leading $m$ eigenvectors and eigenvalues of $ K $. Direct comparison of eigenvectors are difficult, because they are only uniquely defined up to a sign change. Let $ \triangle(\mathcal{U}, \mathcal{V}) $ denote the distance between two subspaces $ \mathcal{U}$ and $ \mathcal{V} $. Here we follow \cite{homrighausen2016nystrom} and define 
\begin{equation*}
	\triangle(\mathcal{U}, \mathcal{V}) =  \mid\mid \Pi_{\mathcal{U}} - \Pi_{\mathcal{V}} \mid\mid_F,
\end{equation*} where $\Pi_{\mathcal{U}}$ and $\Pi_{\mathcal{V}}$ are the orthogonal projection associated with $ \mathcal{U} $ and $ \mathcal{V} $, respectively. The smaller the distance between the subspaces generated by the approximated eigenvectors $U_m$ and the true eigenvectors $V_m$, the better the approximation. To measure the approximation performance of the eigenvalues, for any vector $\hat{\bs{\lambda}}$ containing the estimated leading eigenvalues in descending order, we compare it to the true leading eigenvalues $ \bs{\lambda} $ using $ \mid\mid\hat{\bs{\lambda}} - \bs{\lambda}\mid\mid_{l_2} $. In our numerical experiment, we simulate 1000 random locations in the unit domain. Based on these data points, we compute the correlation matrices K using the Mat\'ern covariance function with $\nu=0.5, \phi=0.167$ and $\nu=2.5, \phi= 0.189$; these correspond to an effective range (defined as the distance at which spatial correlation drops to 0.05) of 0.5 in the unit square spatial domain. A more detailed comparison for different smoothness and effective range is included in the online supplementary materials S.1. Figure \ref{fig:aproxperformance} shows the approximation results for the first 100 eigencomponents when the projection matrix $ \Phi$ is $ [I_{k\times k},0_{k\times (n-k)}]^T $ with rows permuted, or $ K^\alpha\Omega$ with $ \alpha=0,1,2 $. We see that when random projections are used the approximation is improved. In addition, there are advantages in taking $ \Phi$ to be $ K^\alpha\Omega$, where in practice $ \alpha = 1 $ appears to be a good choice. 

\subsection{Random projection for spatial linear mixed models}

Here we illustrate the random projection approach for a spatial linear mixed model (SLMM) with an emphasis on dealing with confounding. \citet{A.Banerjee2012} proposes using random projection for efficient Gaussian process regression. In this subsection we extend their approach so it applies to both SLMMs and the linear restricted spatial regression model. This description also serves as an introduction to our more general approach to SGLMMs. 

For the linear case, model fitting is based on the marginal distribution of $\bs{Y}\mid\bs{\beta},\phi,\sigma^2,\tau^2$. The main computational challenge is therefore due to the expense in calculating inverses and determinants for large covariance matrices. Random projection may be used to approximate the correlation matrix using its principal components. To fit the full model with random projection (FRP), we apply Algorithm \ref{RPalgorithm} to approximate $ R(\phi) \approx \widetilde{R}(\phi) = U_mD_mU_m^T$. We rewrite the model as follows, 
 \begin{equation}\label{eqn:lmrp}
		\begin{aligned}
			\bs{Y} &= X\bs{\beta} + U_mD_m^{1/2}\bs{\delta } +\bs{\epsilon}, \hspace{5mm} \bs{\epsilon} \sim N(\bs{0},\tau^2 {I}), \hspace{5mm} \bs{\delta }\sim N(\bs{0},\sigma^2I).\\
			& \text{Marginally:  }  \bs{Y}\mid\bs{\beta},\phi,\sigma^2,
			\tau^2 \sim N\left( X\bs{\beta},\sigma^2\widetilde{R}(\phi) +\tau^2 I\right).
		\end{aligned}
\end{equation} 
Analogously, our RSR model with random projection (RRP) is
 \begin{equation}\label{eqn:lmrsrrp}
		\begin{aligned}
			\bs{Y} &= X\bs{\beta} + P_{[X]}^{\perp}U_mD_m^{1/2}\bs{\delta } +\bs{\epsilon}, \hspace{5mm} \bs{\epsilon} \sim N(\bs{0},\tau^2 {I}), \hspace{5mm} \bs{\delta }\sim N(\bs{0},\sigma^2I).\\
			& \text{Marginally:   }  \bs{Y}\mid\bs{\beta},\phi,\sigma^2,
			\tau^2 \sim N\left( X\bs{\beta}, \sigma^2{P_{[X]}^{\perp}}\widetilde{R}(\phi){P_{[X]}^{\perp}} +\tau^2 {I} \right).
		\end{aligned} 
\end{equation} 

Hereafter, $\widetilde{R}(\phi)$ will be referred to as $\widetilde{R}$ to suppress its dependency on the unknown parameter $\phi$. Fitting the FRP model (\ref{eqn:lmrp}) involves evaluating the inverse and determinant of $\sigma^2\widetilde{R} +\tau^2 {I} = \sigma^2{U_m}{D_m}{U_m}^T + \tau^2 {I} $. Then, by the Sherman-Morrison-Woodbury identity \citep{algebrabook}, we have $ (\sigma^2{U_m}{D_m}{U_m}^T + \tau^2 {I})^{-1} =  \tau^{-2} {I} - \tau^{-2}{U_m}(\sigma^2{D_m}^{-1}+  \tau^{-2}{U_m^TU_m})^{-1} {U_m}^T \tau^{-2} $. The matrix inversion of $\sigma^2{D_m}^{-1}+  \tau^{-2}{U_m^TU_m}$ can be further reduced to inverting an $ m \times m $ diagonal matrix $\sigma^2{D_m}^{-1} + \tau^{-2}{I} $ with cost of $ m $ flops, since $ U_m$ has orthonormal columns. The determinant calculation can also be simplified. By the determinant lemma \citep{algebrabook},  $|\sigma^2\widetilde{R} +\tau^2{I} | = | \sigma^2{D_m}^{-1}+  \tau^{-2}{I} |\times|\tau^2 {I}|\times|{D_m}|$ = $ \prod (\sigma^2D_{m,ii}^{-1} + \tau^{-2}) \times \tau^{2n}\times\prod D_{m,ii}$. Similarly, fitting the RRP model (\ref{eqn:lmrsrrp}) involves calculating the inverse and determinant of $\sigma^2{P_{[X]}^{\perp}}\widetilde{R}{P_{[X]}^{\perp}} +\tau^2 {I}$, for which the dominant cost is tied to the $ m \times m $ matrix $ \sigma^2{D_m}^{-1}+  \tau^{-2}{U_m^TP_{[X]}^{\perp}U_m} $. 

For the linear case we can simply approximate the correlation matrix of the random effect $R$ with $\widetilde{R}$ without explicitly reparameterizing the random effects. However, for SGLMMs, we do not have closed-form marginal distribution. It is therefore necessary to obtain the reduce random effects $ \bs{\delta } $ and carry out inference based on $ \pi( \bs{\theta,\beta,\delta \mid Z}) $.

\subsection{Random projection for spatial GLMMs}

Here we describe how to reparameterize and reduce the dimension of our models such that the resulting model is easier to fit and preserves the desirable properties of the original model. We do this for both cases, first where confounding may not be an issue and the second where we want to address confounding. 

We apply Algorithm \ref{RPalgorithm} to $R$ to obtain $U_m$ and $D_m$. If confounding is not an issue, we replace random effect $ \bs{W} $ with $U_m D_m^{1/2}\bs{\delta}$. The SGLMM \eqref{eqn:sglmm} may be rewritten as 
 \begin{equation}\label{eqn:FRP}
		\begin{aligned}
			g\left\lbrace E(Z_i|\bs{\beta},U_m, D_m,\bs{\delta })\right\rbrace & = X_i\bs{\beta} + (U_mD_m^{1/2})_i \bs{\delta }, \\
			\bs{\delta }\mid\bs{\theta} & \stackrel{approx}{\sim}  \text{N}(\bs{0},\sigma^2I).
		\end{aligned}
\end{equation} 
We refer to this as the full model with random projection (FRP). Essentially, the spatial dependence in $ \bs{W} $ is transformed into a reduced-dimension spatially independent variable $\bs{\delta }$ and synthetic spatial variable $U_mD_m^{1/2}$. This combines the idea of spatial filtering \citep{GetisGriffith} and PCA, thereby reducing the dimension of the posterior distribution. By also reducing the correlations among the parameters, our approach improves the mixing of MCMC algorithms for sampling from the posterior. Once priors $ p(\bs{\beta}, \bs{\theta}) $ are specified, we can sample from the full conditionals (see the online supplementary materials S.2) using Metropolis-Hastings random-walk updates. 

To address the confounding problem, we follow the RSR approach to restrict the random effects to be orthogonal to the fixed effects \citep{Hodges2010}. We can project our reduced random effects $U_mD_m^{1/2}\bs{\delta}$ to the orthogonal span of $X$. The restricted model with random projection (RRP) can be summarized as follows:
 \begin{equation}\label{eqn:RRP}
		\begin{aligned}
			g\left\lbrace E(Z_i\mid\bs{\beta},U_m,D_m,\bs{\delta })\right\rbrace 
			& =  X_i\bs{\beta} + (P_{[X]}^{\perp} U_mD_m^{1/2})_i\bs{\delta },\\
			\bs{\delta }\mid\bs{\theta} &\stackrel{approx}{\sim} \text{N}(\bs{0},\sigma^2I).
		\end{aligned}
\end{equation} 
Fitting RRP is similar to FRP, except that in the data likelihood $ \prod_{i=1}^n f\left( Z_i\mid\bs{\beta},U_m,D_m,\bs{\delta }\right) $, $U_m$ is replaced by $ P_{[X]}^{\perp}U_m $.

It is tempting to first replace $ \bs{W} $ by $P_{[X]}^{\perp}\bs{W} \sim \text{MVN}(\bs{0}, \sigma^2P_{[X]}^{\perp}RP_{[X]}^{\perp})$, then reduce the dimension of random effects by approximating the matrix $ P_{[X]}^{\perp}RP_{[X]}^{\perp} $. However the order of approximation and projection affects the inference of the covariance parameter $ \phi $. Although the projected eigenvectors of $R$, i.e. $ P_{[X]}^{\perp} U_m $, is the same as the eigenvectors of $ P_{[X]}^{\perp}RP_{[X]}^{\perp} $, the ordering of the eigencomponents may change depending on the direction of projection. To better approximate the original random effects $ \bs{W} $, rather than $P_{[X]}^{\perp}\bs{W}$, we therefore first approximate the correlation matrix and then perform the requisite orthogonal projection. 

\subsection{Random projection for areal data}\label{sec:areal}

Our approach reduces the dimension by decomposing its correlation matrix. Hence, it can be easily applied to Gaussian Markov random field models as well. Here we develop FRP and RRP for non-Gaussian areal data. Note that RRP model for areal data is similar to the approach proposed by \citet{HughesHaran}. Both methods adjust for confounding and reduce the dimension of random effects. The only difference is that we decompose the covariance matrix, whereas their decomposition is performed on the Moran operator (details are provided later in this subsection).

Consider spatial data located on a discrete domain, for instance mortality rates by county across the U.S. If we describe the data locations via nodes on an undirected graph with edges only between nodes that are considered neighbors, we can model the spatial dependence via a Gaussian Markov random field model. To model the dependence of $ \bs{W}=\left( W_1,...,W_n \right)^T$, where the index indicates block, we define the neighboring structure among blocks through an $n \times n$ adjacency matrix $ A $ with $\text{diag}(A) = 0$ and $A_{ij} = 1$ if the $i^{th}$ and $j^{th}$ locations are connected, or $A_{ij}=0$ if they are not connected \citep{Besag1991}. A common model for $\bs{W}$ is an intrinsic conditionally auto-regressive (ICAR) or Gaussian Markov Random Field prior: 
 \begin{equation}\nonumber
		p(\bs{W}\mid\tau) \propto \tau^{rank (Q)/2} \exp \left( -\frac{\tau}{2} \bs{W}^TQ\bs{W} \right), 
\end{equation} 
where $\tau$ is a smoothing parameter that controls the smoothness of the spatial field, and $Q = \text{diag}(A\bs{1}) -A$ is the precision matrix ($ \bs{1} $ is a $n$-dimensional vector of 1s).

For the discrete domain, \citet{HughesHaran} use the eigenvectors of the Moran operator, $M = P_{[X]}^{\perp}AP_{[X]}^{\perp}$, to reduce the dimension of the random effects.
Their method alleviates confounding while preserving spatial dependence structure implied by the underlying graph. The eigenvectors can be interpreted as spatial patterns corresponding to different degrees of spatial dependency. 

We can also fit both FRP and RRP to areal data and achieve similar dimension reduction and computational gains. First we obtain the covariance matrix by taking the generalized inverse of the precision matrix $ Q $. Then apply random projection on the covariance matrix $ Q^{-1} $, and proceed with either FRP or RRP as described in the preceding section. The eigenvectors corresponding to large eigenvalues represent large-scale spatial variation. The advantage of this PCA approach is that a relative small number of PC's is enough to capture most spatial variation. Computationally, our method is not as efficient as \cite{HughesHaran} because of the extra cost of inverting the precision matrix. However, for the ICAR model, the precision matrix is fixed and defined beforehand, so the inversion and random projection only need to be performed once. The model estimates from RRP are comparable to those obtained by \cite{HughesHaran} (see the online supplementary materials S.4). 

\subsection{Rank selection}\label{sec:rank}
Here we provide a general guideline for selecting the rank for projection-based models. From a Bayesian perspective, the rank can be determined by model comparison criteria such as DIC \citep{DIC}. We can fit several models using different number of ranks and select the one with the smallest DIC. However, to reduce computational time, we recommend the following procedure to select the appropriate rank before fitting either FRP or RRP model. 

Since the projection-based models combine the idea of spatial filtering and PCA dimension reduction, we can fit non-spatial generalized linear models with predictors $ X $ and synthetic spatial variables $ U_mD_m^{1/2} $ for $ m = 1,2\dots $, and then select the initial rank based on variable selection criterion such as BIC. The synthetic spatial variables $ U_mD_m^{1/2} $ can be obtained by performing eigendecomposition on the correlation matrix $ R^{(0)} $ for an appropriate range value $ \phi^{(0)} $, for example, $ \phi^{(0)} $ is half of the maximum distance among observations. If performing a full eigendecomposition is computationally infeasible, one may perform random projection using Algorithm \ref{RPalgorithm} to approximate the leading eigencomponents. In our binary spatial data simulation study, the overall smallest BIC corresponds to $ m=75 $ for $ \nu = 0.5 $, $ m=50 $ for $ \nu = 1.5 $, $ m=40 $ for $ \nu = 2.5 $, and $ m=30 $ for $ \nu = \infty $. To be careful, we then advocate seeing if increasing the rank leads to a marked improvement in the model, for instance by again using criteria like DIC. If it does, it may be useful to try increasing the rank again; if there is not much change, we can stop and simply use the current rank. For each of the above smoothness values, we also studied multiple simulated examples using several ranks to study the effectiveness of the rank selection method and the performance of our projection-based approach. Our simple heuristic appears to work well -- the DIC values from model fits agree with BIC selection, and the prediction performance increases by only a small margin above the selected rank. We note that in one particular case, the Poisson SGLMM with a very rough latent process ($\nu<2$), we found that DIC suggests much larger ranks than perhaps necessary. If researchers want a more rigorous  comparison of models with different ranks and are willing to implement Bayes factor calculations, comparing Bayes factors would be an   alternative to what we have proposed.



\subsection{Computational gains}
The advantages of our reparameterization schemes are shorter computational time per iteration and less MCMC iterations to achieve convergence. These result from: (1) reducing large matrix operations, (2) reducing the number of random effects, and (3) improving MCMC algorithm mixing. Although the main computational cost of our approach is of order $ O(n^2m) $ from applying Algorithm \ref{RPalgorithm}, it is dominated by matrix multiplications that can be easily parallelized by multi-core processors. Leveraging parallel computing for matrix multiplication, the remaining dominant cost of fitting our model is of order $ O(nm^2) $ due to the singular value decomposition of $ n\times m $ matrices. To illustrate the computational gain of the projection-based models, we fit both the SGLMM and the projection-based models to simulated Poisson data. We fit the SGLMM using one-variable-at-a-time Metropolis-Hastings random-walk updates. To fit the projection-based models, we update random effects $ \bs{\delta } $ in a block using spherical normal proposal; simple updating scheme for $ \bs{\delta } $ is sufficient because it has a smaller dimension and are decorrelated. In our implementation, Algorithm \ref{RPalgorithm} is coded in C++ using Intel' Math Kernel Library BLAS and LAPACK routines for matrix operations; the MCMC is written in programming language R \citep{CiteR}. All the code was run on National Center for Atmospheric Research's Yellowstone supercomputer \citep{yellowstone}.

\begin{figure}[hpt] 
	\centering
	\includegraphics[scale=0.7]{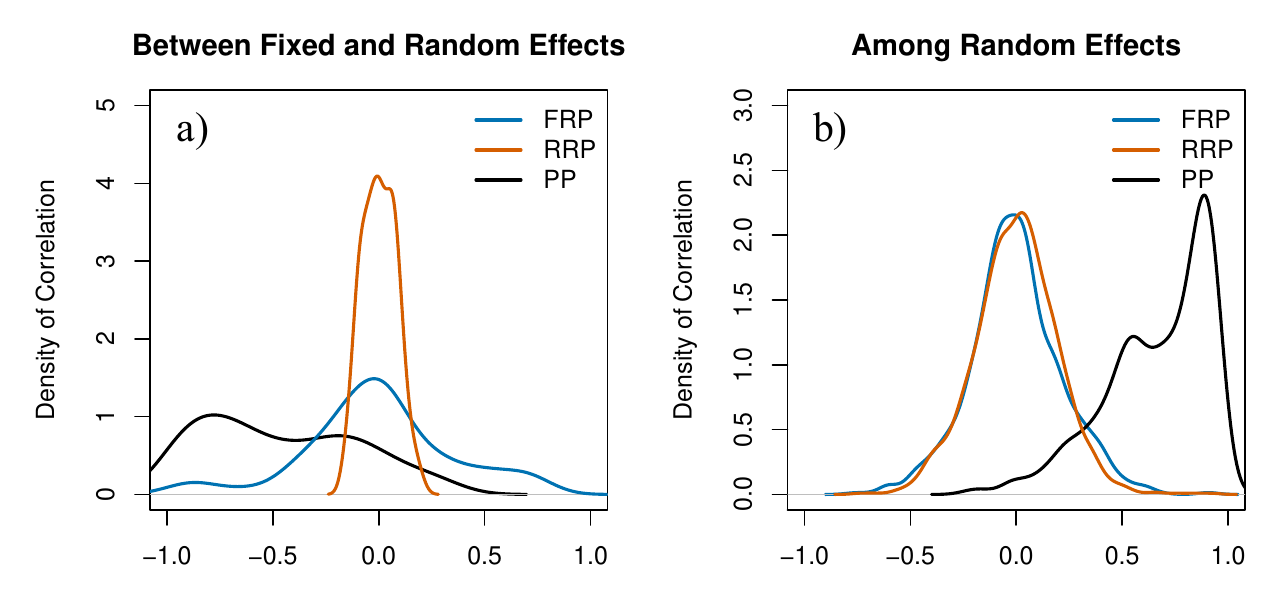}\\
	\caption{Pairwise cross-correlations are close to zero for both projection-based approaches (FRP and RRP). This is true for each pair of random effects (right) and between fixed and random effects (left). In contrast, the cross-correlations for the predictive process (black curves) are much larger. Hence our approaches result in better MCMC mixing. FRP = full model with random projection, RRP = restricted model with random projection and PP = predictive process model.}
	\label{fig:cor}
\end{figure}

\begin{figure}[hpt]
	\centering
	\includegraphics[scale=0.4]{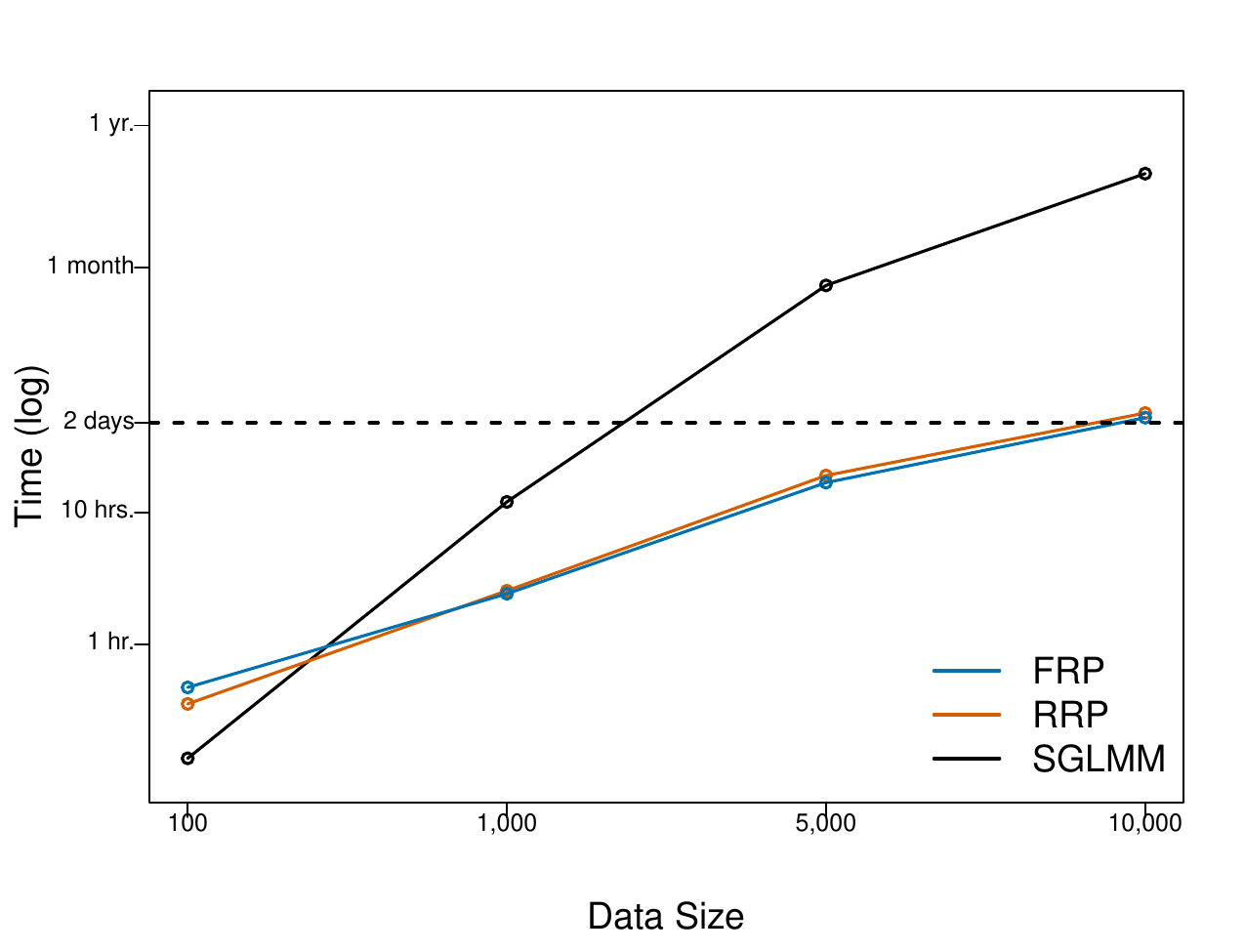}
	\caption{Computational time for $ 10^5 $ iterations versus data size for SGLMM and projection-based approach (both FRP and RRP) with rank 50. This illustrates the benefit of the projection-based approach (both FRP and RRP) in terms of computational time. FRP = full model with random projection, and RRP = restricted model with random projection.} 
	\label{fig:time}
\end{figure}

To see the improvement in MCMC mixing, we compute the effective sample size (ESS) using the R \textit{coda} package \citep{coda}; it provides the number of independent samples roughly comparable to the number of dependent samples produced by the MCMC algorithm, therefore a larger ESS implies better Markov chain mixing. Based on our results, the projection-based models have better mixing, for example, the univariate ESSs of the RRP are, on average, 12 times larger than the ones from the SGLMM and three times the ones from the predictive process \citep[using the R package by][]{finley2013spbayes} for the same number of MCMC iterations. The mixing improvement of our projection-based models is implied from the \textit{a posteriori} correlations (Figure \ref{fig:cor}); our projection-based models (both FRP and RRP) produce weakly correlated random effects compared to the predictive process. The improvement in computational time is illustrated in Figure \ref{fig:time}. The time required increases dramatically for SGLMM as the data size increases, however we can still fit the random projection model in a reasonable amount of time. We also compute ESS per second to compare MCMC efficiency; our RRP model more than 120 times more efficient than the SGLMM.  


\section{Simulation Study and Results}\label{sec:simu}
In this section, we apply our approaches to simulated linear, binary and Poisson data. For each case, we simulate 100 data sets where the locations are in the unit domain $ [0,1]^2 $. We fit both FRP and RRP models to simulated data with size of $ n=1000 $ at random locations, then make predictions on a $ 20\times 20 $ grid. We adjust the regression parameters of RRP using equation \eqref{eqn:adjust} (denote this adjusted inference A-RRP). Throughout the simulation study, we let $ X $ be the xy-coordinate of the observations and $ \bs{\beta} = (1,1)^T $. We simulate $ \bs{W} $ from the Mat\'{e}rn covariance function with $ \bs{\theta} = (\nu, \sigma^2, \phi)^T = ( 2.5, 1, 0.2)^T $, which has the form as below \citep[Section 4.2]{RasmussenGPM}: 
 \begin{equation*}
		C(h) = \sigma^2  \left(1 +\frac{\sqrt{5}|h|}{\phi} + \frac{5|h|^2}{3\phi^2} \right) \exp\left(-\frac{\sqrt{5}|h|}{\phi}\right)
\end{equation*} 

We use a vague multivariate normal prior $ N(\bs{0}, 100I) $ for regression coefficients $ \bs{\beta} $, inverse gamma prior $ IVG(2, 2) $ for $ \sigma^2 $ and uniform prior $ U(0.01, 1.5) $ for $ \phi $. We have experimented with different choice of prior; the inference performances are similar. To evaluate our approaches, we compare inference performance with a focus on the posterior mean estimates and 95\% equal-tail credible intervals of $ \bs{\beta} $, and we compare prediction performance based on mean square error.

\subsection{Linear case}
The random projection models are first assessed under the linear case (details are presented in the online supplementary materials S.3). Let $ \bs{x}_1$, $ \bs{x}_2 $ denote the xy-coordinates. We simulate data from
 \begin{equation*}
		\bs{Y} = \bs{x}_1 + \bs{x}_2 + W +\epsilon, \hspace{5mm} \bs{\epsilon} \sim N(\bs{0},\tau^2 {I}), \hspace{5mm} \bs{W} \sim \text{MVN}(\bs{0},\sigma^2{R}(\phi)),
\end{equation*} 
where the noise $\bs{\epsilon}$ has variance ${ \tau^2 = 0.1 }$. 

We fit both FRP and RRP models using rank $ m = 50 $ based on the marginal distribution of ${\bs{Y}\mid\bs{\beta},\phi,\sigma^2,\tau^2}$ as described in \eqref{eqn:lmrp} and \eqref{eqn:lmrsrrp}, and we use $IVG(2,1)$ prior for $\tau^2$. For the linear case, fitting the full SLMM and RSR model is fast for data of size $ {n = 400} $, so we compare results across all four models. Our results show that inference and prediction provided by the random projection models are similar to the original models they approximate. As noted by \citet{ephraim2015}, when the data are simulated from the full SLMM, we see a low $ \bs{\beta} $ coverage for the RSR model; therefore, its approximated version RRP also has a low coverage. However, this problem is resolved after a simple adjustment (A-RRP) as recommended by \cite{ephraim2015}.

We also conduct a simulation study for larger data size $ {n = 1000} $, and we fit both FRP and RRP with rank $ m = 50 $. Our results show that the distributions of $ \bs{\beta} $ estimates for both FRP and RRP are centered around the true value, and the distributions are comparable. Coverage of 95\% credible intervals for FRP and A-RRP are comparable to the nominal rate. For prediction performance, the mean square error is similar for both models and the predicted observations at testing locations recover the spatial patterns well.

\subsection{Binary case}
The main goal of our approximation method is to fit spatial generalized linear mixed models for large data sets. Here we examine our model performance under the binary case generated with a logit link function $ \text{logit}(p) = \log\left\lbrace p/(1-p) \right\rbrace $. We compare two simulation schemes: the confounded case $ \bs{\eta}= \bs{x_1} +\bs{x_2} + \bs{W}$, and the orthogonal case $\bs{\eta}= \bs{x_1} +\bs{x_2} + P_{[X]}^{\perp}\bs{W}$. For both cases we use the same parameter values as the linear case and simulate $\bs{W}$ from $N(\bs{0},\sigma^2R(\phi))$. We consider two simulation schemes because in practice we do not know whether there are spatial latent variables that may be collinear with our covariates. A careful approach, therefore, involves fitting both FRP \eqref{eqn:FRP} and RRP \eqref{eqn:RRP} models under both schemes to get a fair assessment of the FRP and RRP approaches. Because it is hard to fit full SGLMMs and the RSR models for moderate data size, we compare only FRP and RRP models for 100 simulated datasets of size $ n = 1000 $. Although we do not have a comparison with the original model fit, we can look at how well the true parameters are recovered and compare the prediction mean square error to judge the projection-based models. 

\begin{figure}[hpt]
	\centering
	\includegraphics[scale=0.5]{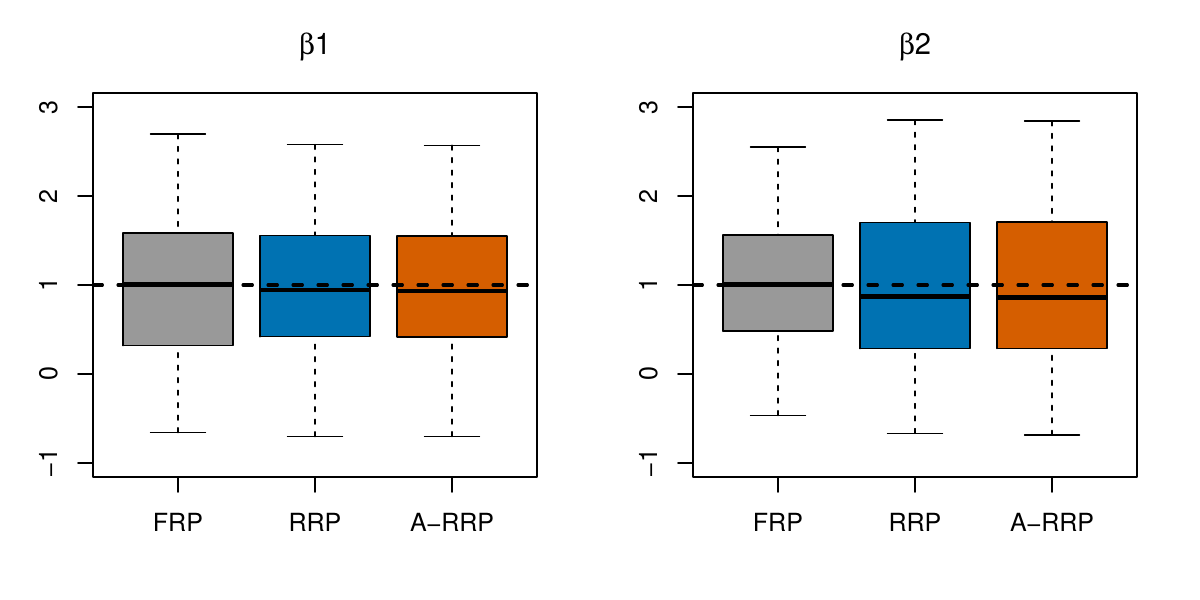} \\
	\includegraphics[scale=0.5]{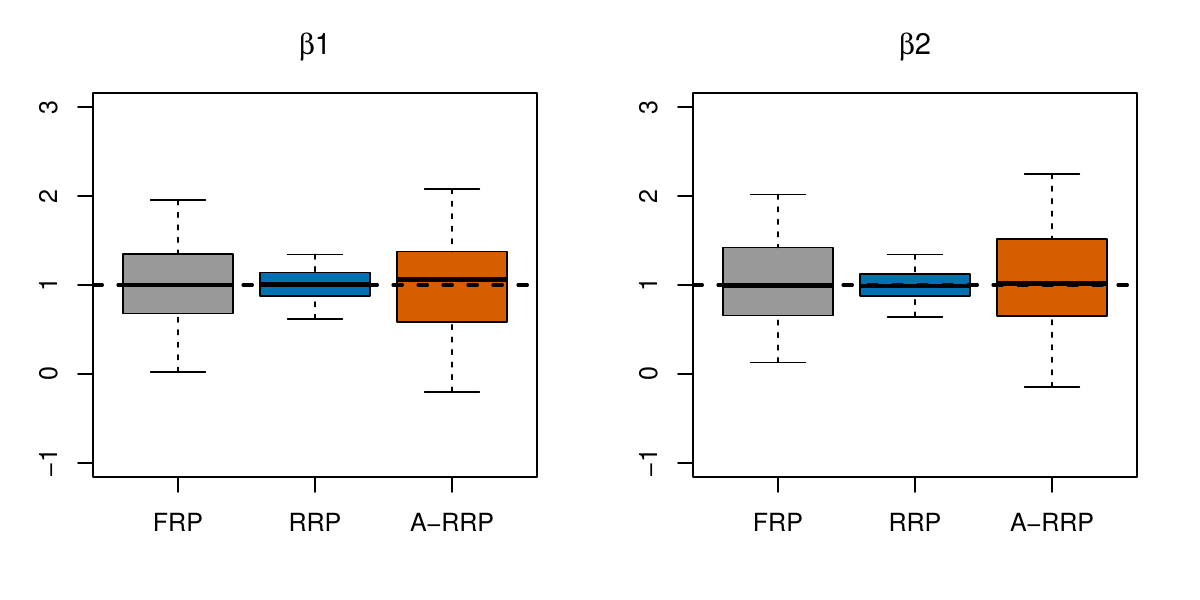}            
	\caption{Binary simulation study: distribution of $ \bs{\beta} $ posterior mean estimates for RP models and after adjustment. First row for the confounded case, and second row for the orthogonal case. All distributions center around the true value. For the confounded case (top row), FRP and RRP have similar results; while under the orthogonal case (bottom row), RRP produce much tighter distribution. For both cases, the adjusted inference A-RRP is similar to the FRP. FRP = full model with random projection, RRP = restricted model with random projection, A-RRP = adjusted inference for RRP.}
	\label{fig:InferenceBetaBin1000}
\end{figure}

\begin{figure}[hpt]
	\centering
	\includegraphics[scale = 1.2]{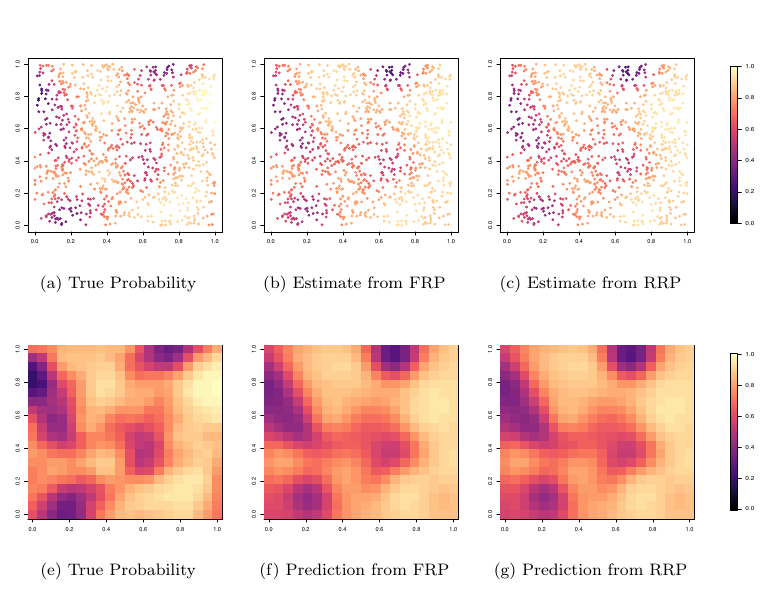}   
	\caption{First row shows the estimated probability surface at all training locations. Second row shows the predicted probability surface on a 20x20 grid using random projection models. Left column is simulated data, middle column shows the FRP, and right column shows the RRP. Comparing the patterns from the true model (left column) to the ones from projection-based models using rank $ m=50 $ (middle and right columns), we see the projection-based models are able to recover the true value quite well. FRP = full model with random projection, RRP = restricted model with random projection, A-RRP = adjusted inference for RRP.}
	\label{fig:Bin_pred1000}
\end{figure}

Our simulation results show that under the confounded case, $ \bs{\beta} $ estimates for both FRP and RRP have similar distributions (Figure \ref{fig:InferenceBetaBin1000}). However the coverage of RRP, about 41\% , is much lower than the 95\% nominal rate. This is because the credible intervals obtained under the RRP are similar to the ones for RSR models, which are likely to be inappropriately narrow \citep{ephraim2015}; the mean length of the credible intervals (with 95\% intervals) under RRP model is 0.84(0.70, 1.09) compared to 4.15(2.61, 7.01) under the FRP model. However this problem is resolved after the adjustment; the coverage of A-RRP is comparable to the nominal rate and its interval length is 4.17(2.94,7.08), similar to the one from the FRP model. Under the orthogonal case, in contrast, RRP performs much better than FRP. The point estimates from RRP are distributed tightly around the true values (Figure \ref{fig:InferenceBetaBin1000}). Its credible interval has better coverage than the FRP; they are 94.95\% and 100\% respectively. Moreover, RRP has much narrower credible intervals, it is 0.81(0.733, 0.95) compared to 4.11(2.75, 6.66) under the FRP. And again the adjusted inference A-RRP is similar to that of FRP. Figure \ref{fig:Bin_pred1000} shows the estimated probability surface for the binary field at the training locations and the predicted probability surface at the testing locations under the confounded simulation scheme. We see that our projection-based approaches work well in recovering the true spatial pattern (results for the orthogonal simulation scheme are similar, hence not shown). Although the predictive surface seems somewhat smoother than the true surface, this could be because binary outcomes do not provide enough information for the latent variable.


\subsection{Poisson case}
\begin{figure}[hpt]
	\centering
	\includegraphics[scale = 1.2]{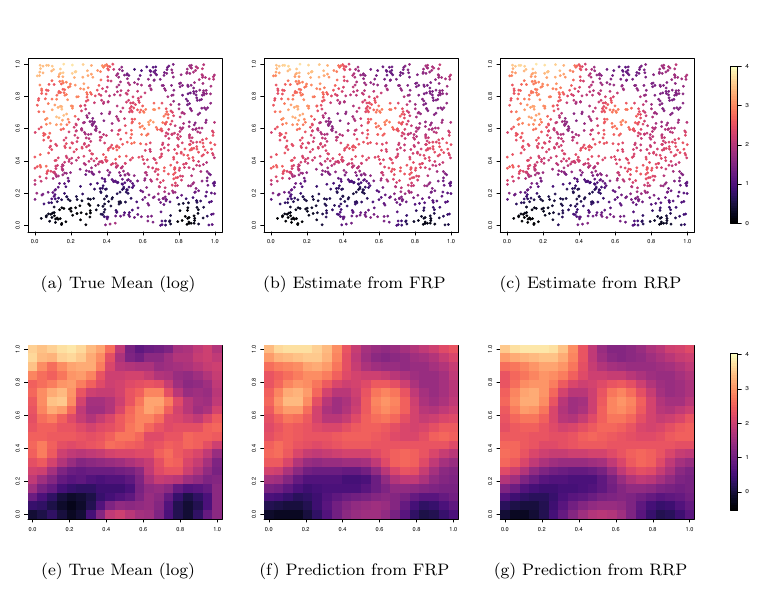} 
	\caption{Simulated Poisson example: First row shows the estimated Poisson mean surface (in log scale) at all training locations. Second row shows the predicted Poisson mean surface (in log scale) on a 20x20 grid using projection-based models. Left column is simulated data, middle column shows the FRP, and right column shows the RRP. Comparing the pattern from the true model (left column) to the ones from our models using rank $ m=50 $ (middle and right columns), we see the projection models are able to recover the true value quite well. FRP = full model with random projection, RRP = restricted model with random projection, A-RRP = adjusted inference for RRP.}
	\label{fig:Poi_pred1000}
\end{figure}

We also examine our model performance under the Poisson case. The results are similar to the Binary case; hence, for brevity we summarize our results here and present the full results in the online supplementary materials. We simulate Poisson data with a natural logarithm link function using the same parameter values as the linear case; again, both simulation schemes are considered. Under the confounded simulation scheme, FRP and RRP have similar distribution for point estimates; RRP provides precise but inaccurate estimates, but after adjustment, A-RRP produces reasonable coverage. Under the orthogonal simulation scheme, RRP performs much better than FRP in terms of both point estimates and credible intervals. For both cases, the adjusted inference A-RRP is similar to the FRP, hence we can fit only the RRP model for its computational benefits and recover the results for fitting FRP. Figure \ref{fig:Poi_pred1000} shows the estimated expectation of the Poisson process (log scale) at the training locations and the predicted expectation (log scale) at the testing locations under the confounded simulation scheme. We see that the projection-based models work well in recovering the true (results for the orthogonal simulation scheme are similar, hence not shown).



\subsection{Comparison to Predictive Processes}
The predictive process approach \citep{Banerjee2008} has been very influential among reduced rank approaches. In the context of spatial generalized linear mixed models for non-Gaussian data, we believe our approach offers some benefits over the predictive process approach: (i) We avoid having to choose the number and locations of knots. Instead, our approach requires specifying the rank for which we have a heuristic; no further user specifications are required. (ii) We provide an approach to easily alleviate spatial confounding. (iii) The reparameterization in our projection-based approach results in decorrelated parameters in the posterior, thereby allowing for a faster mixing MCMC algorithm. Simulation results show that our approach is comparable to the predictive process in terms of inference and better in terms of prediction. Our approach also allows us to fit and study both the restricted and non-restricted versions of the SGLMM. Results for the comparison are shown in the online supplementary materials S.5.  

\section{Applications}\label{sec:application}
\subsection{Binary data application}
We apply our approach to classify rock types using a reference synthetic seismic data set. Fluvsim is a computer program that produces realistic geological structures using a sequential scheme; it is used for modeling complex fluvial reservoirs \citep{fluvsim}. The high-resolution 100x120x10 3-dimensional grid data set is simulated from the program fluvsim conditioning on well observations. A similar reference data set obtained from fluvsim has been used to test the classification method in \cite{well2008}. Here we illustrate our projection-based approach on one layer of the rock profile. In the data set, there are five rock types: crevasse, levies, border, channel and mud stone. We combine crevasse and mud stone as one group and treat the rest as the other group for binary classification. Along with the rock type data, we also have acoustic impedance data that is associated with the rock properties; it is desirable to identify rock types from seismic-amplitude data using statistical methods \citep{well2008}. 

We fit both FRP and RRP models at 2000 randomly-selected locations and predict the rock profile on a 24x30 grid. Prior to fitting the projection-based models, the rank is selected by fitting non-spatial logistic regression models with synthetic spatial variables as described in Section \ref{sec:rank}. The BIC values from the resulting models suggest that rank $ m=50 $ is sufficient. In order to help diagnose convergence, we ran multiple chains starting at dispersed initial values and compared the resulting marginal distributions while also ensuring that the MCMC standard errors for the expected value of each parameter of the distribution was below a threshold of 0.02 \citep[cf.][]{FlegalHaranJones2008}. 

\begin{figure}[hbt]
	\centering
	\includegraphics[scale=0.35]{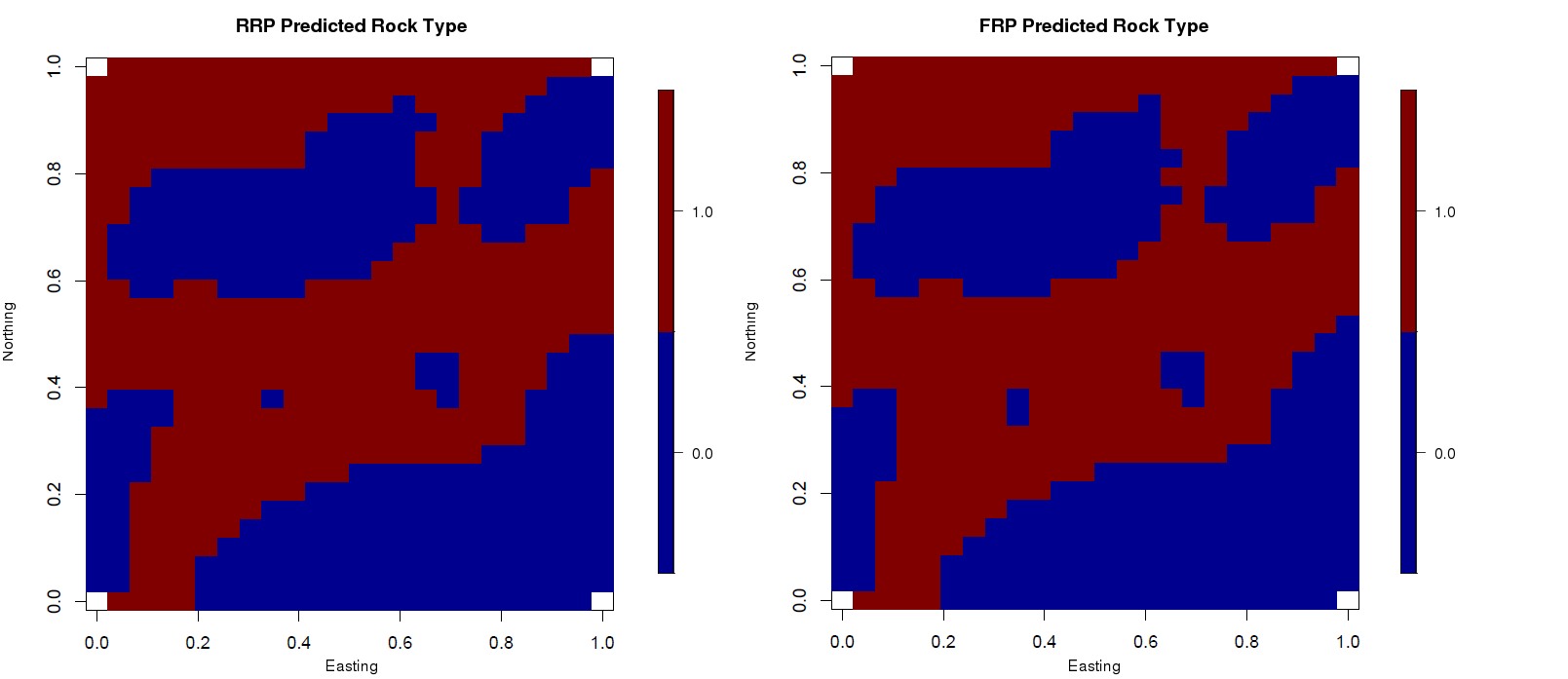}
	\caption{Predicted rock profile by restricted model with random projection (right) and full model with random projection (left) using rank $ m=50 $.}
	\label{fig:wellpred}
\end{figure}

Estimated coefficients corresponding to x-coordinates, y-coordinates and impedance covariates differ between FRP and RRP models, which are $ {(-0.717, 0.448, -0.233)^T}$ and $ (-1.587$, $ 2.435 $, $-0.396)^T $, respectively. Although interpretations for the estimates are slightly different the predictions, which are of primary interest, are identical between the two models (see Figure \ref{fig:wellpred}). We also assess the predicted rock profile using higher ranks; however, the results are similar to using rank 50, hence they are not shown here. The time to fit either the FRP or RRP model is about 10 hours, whereas for the full model, it would have taken about three weeks to run the same number of MCMC iterations. In general, fitting SGLMM to binary observations is harder due to the poor Markov chain mixing; therefore comparison with the full model is prohibitively expensive. 

\subsection{Count data application}
\begin{figure}[hpt!]
	\centering
	\subfloat[Bird Count Observations]{\includegraphics[width=10cm,height=6cm]{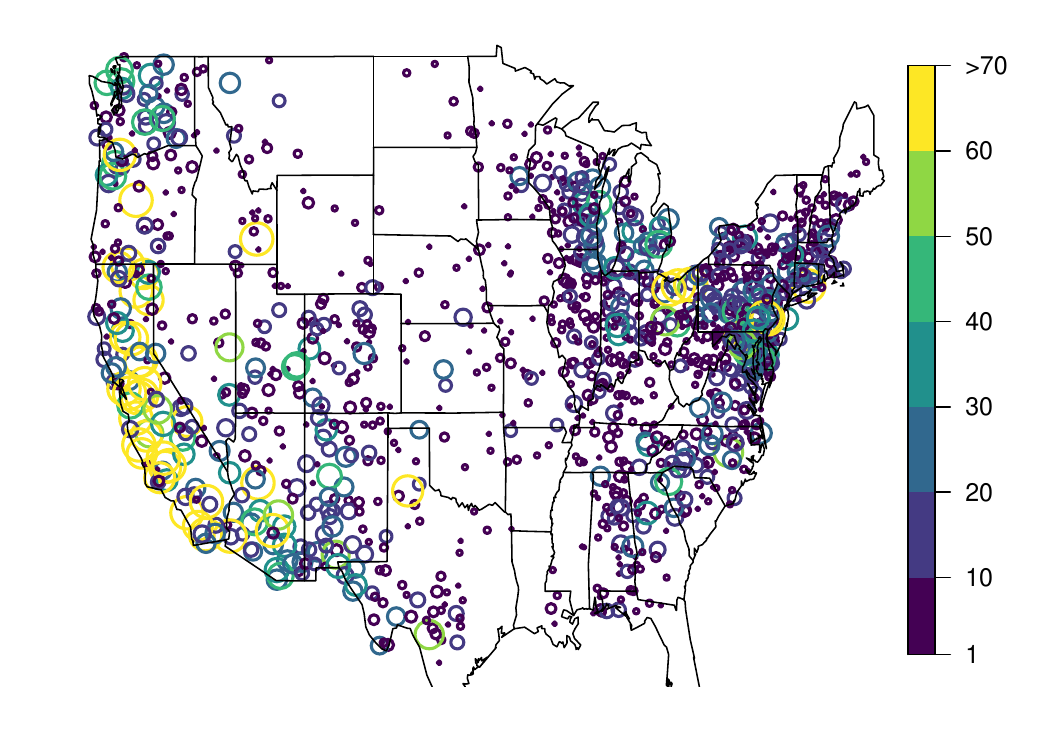}} \hspace{0.5cm}
	\subfloat[Bird Count Prediction]{\includegraphics[width=10cm,height=6cm]{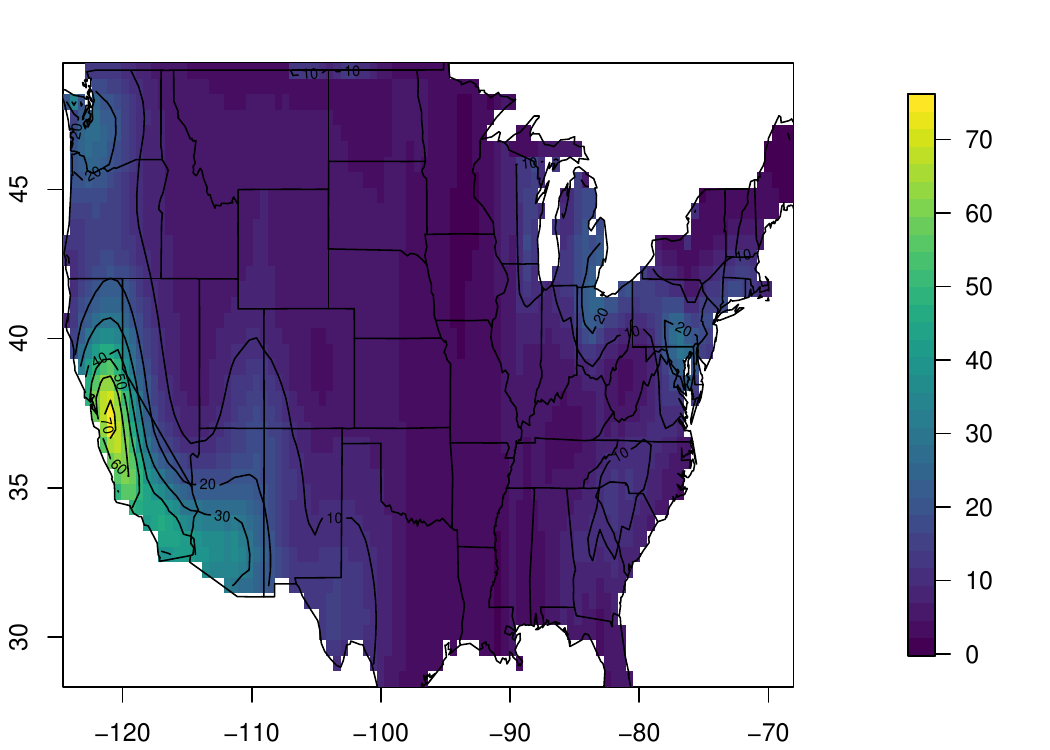}}
	\caption{Data on abundance of house finch in the US from North American Breeding Bird Survey Dataset (a). Size of the circle is proportional to the bird counts. Predicted bird counts on a grid using the random projection method (b). The large number of observed counts on west coast and lack of observations in the central mid-west region is reflected on the prediction map. }
	\label{fig:bbspred}
\end{figure}

Here we illustrate the usefulness of the projection-based models in the context of an environmental study. We consider the relative abundance of house finch (\textit{Carpodacus mexicanus}), a bird species that is native to western North America \citep{elliott1953origin}. Figure \ref{fig:bbspred} shows the number of bird counts obtained in 1999 by the North American Breeding Bird Survey with the size of the circle is proportional to the number of counts. The bird surveys are obtained along more than 3,000 routes across the continental US. There are 50 stops per route, spaced roughly 0.5 miles apart. The observer make a three-minute point count at each stop. The bird count is then the total number of birds heard or seen for all 50 stops \citep{bbsdata}. 

The data set being analyzed has 1257 highly irregular sampling locations. Here we fit the FRP model to approximate the SGLMM with only the intercept term for spatial interpolation. The time to fit FRP is about 7 hours, while the full model would take almost 2 days for the same number of MCMC iterations. Figure \ref{fig:bbspred} shows the abundance map predicted by FRP on a high resolution of 40 x 100 grid. Not surprising, the abundance map is smooth. This reflects that the bird counts are very small in the center and most of the east coast of the US. Our map is also consistent with the observation that large counts are centered near New York area and the West Coast. 

\section{Discussion}\label{sec:discuss}
In this paper, we have proposed projection-based models for fast approximation to SGLMMs and RSR models. Our simulation study shows that our low rank models have good inference and prediction performance. The advantages of our approach include: (1) a reduction in the number of random effects, which lowers the dimensionality of the posterior distribution and decreases the computational cost of likelihood evaluations at each iteration of the MCMC algorithm,; (2) reparameterized and therefore approximately independent random effects, resulting in faster mixing MCMC algorithms; and (3) the ability to adjust for spatial confounding.

Our simulation study shows both the restricted and unrestricted models provide similar results in prediction. RRP provides superior inference when the true model does not have confounding (and hence the spurious confounding effect needs to be removed); it is also computationally more efficient due to its faster mixing. Therefore, we recommend that in general users fit RRP models. If there is concern that the true model may actually exhibit confounding, we recommend adjusting the fixed effects \textit{a posteriori} to recover the inference from FRP as recommended in \cite{ephraim2015}. As we demonstrate here, this is easy to do in practice. We have conducted a simulation study for the Mat\'ern covariance function with $\nu = 0.5, 1.5 $ and several range values. Our study suggests that varying $\phi$ does not affect the results from our projection-based approach. For small values of $\nu$, the projection-based approach still works well in the binary case, offering a large reduction in dimensions without a change in prediction ability. However, in the Poisson case, for non-smooth processes, it may not always be possible to reduce dimensions to the same extent. This finding is consistent with observations by others regarding reduced-rank approaches in the linear Gaussian process setting \citep[cf.][]{Stein20141}.

The current methods rely on parallelization to handle large matrix computations; we have successfully carried these out for $n$ of around 10,000. If we combine parallelization with a discretization of possible values of $\phi$ (to allow for pre-computing the eigendecomposition of the covariance matrix), this approach will likely scale to tens of thousands of data points. The INLA approach \citep{rue2009inla} provides a fast approximate numerical method for carrying out inference for latent Gaussian random field models. An interesting avenue for future research is combining our reduced-dimensional reparameterization with INLA.

There have been a number of recent proposals for dimension reduction and computationally efficient approaches for spatial models. These include the fixed rank approximation by \cite{Cressie2008}, predictive process by \cite{Banerjee2008} and random projection approach for the linear case \cite{A.Banerjee2012}. Our approach can be thought of as a fixed rank approach, but we use the approximated principal eigenfunctions as our basis. The advantage is that we have independent basis coefficients and our approximation minimizes the variance of the truncation error as described in Section \ref{sec:RP}. Our approach is also related to the predictive process in that we effectively subsample random effects \citep[see discussion in][]{A.Banerjee2012}. Developing extension of this methodology to spatial-temporal and multivariate spatial processes may provide fruitful avenues for future research.

%% file: SglmmProjSupp_arxiv.tex
\section*{Supplementary materials to ``A Computationally Efficient Projection-Based Approach for Spatial Generalized Linear Mixed Models" by Guan and Haran}

	\subsection*{S.1 Eigencomponent Approximation Performance}
	
\begin{figure}[hpt]
		\centering
		\includegraphics[page=1]{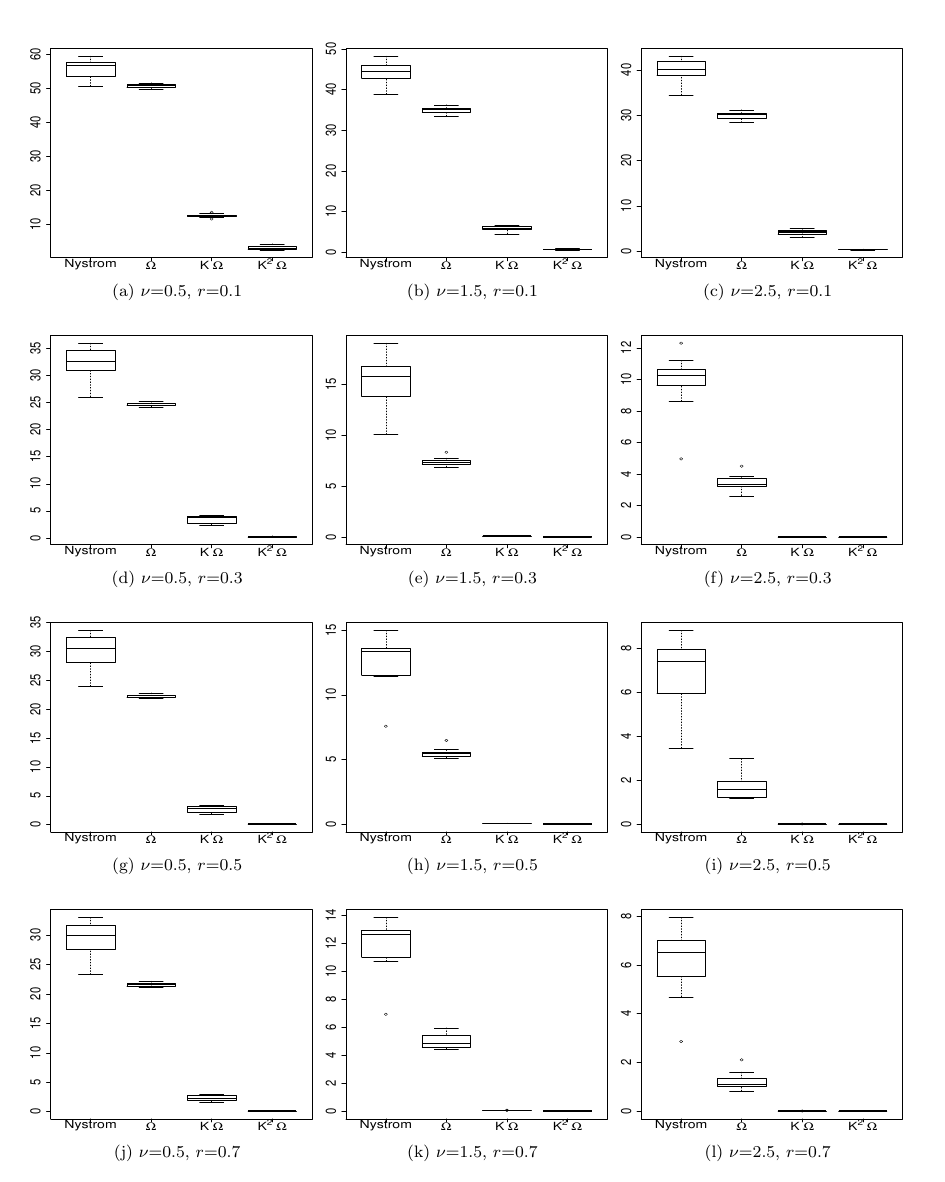}
		\caption{Distance between the subspaces generated by the fist 100 approximated and true eigenvectors under different smoothness $\nu$ and effective range r. Introducing randomness in $\Phi=K^\alpha\Omega$ improves the Nyst\"{o}m approximation to the eigenvectors. Letting $ \Phi $ to be $ K^\alpha\Omega $ with small power $ \alpha =1,2 $ further improves approximation.}
		\label{fig:eigendist}
	\end{figure}

\begin{figure}[hpt]
		\centering
		\includegraphics[page=2]{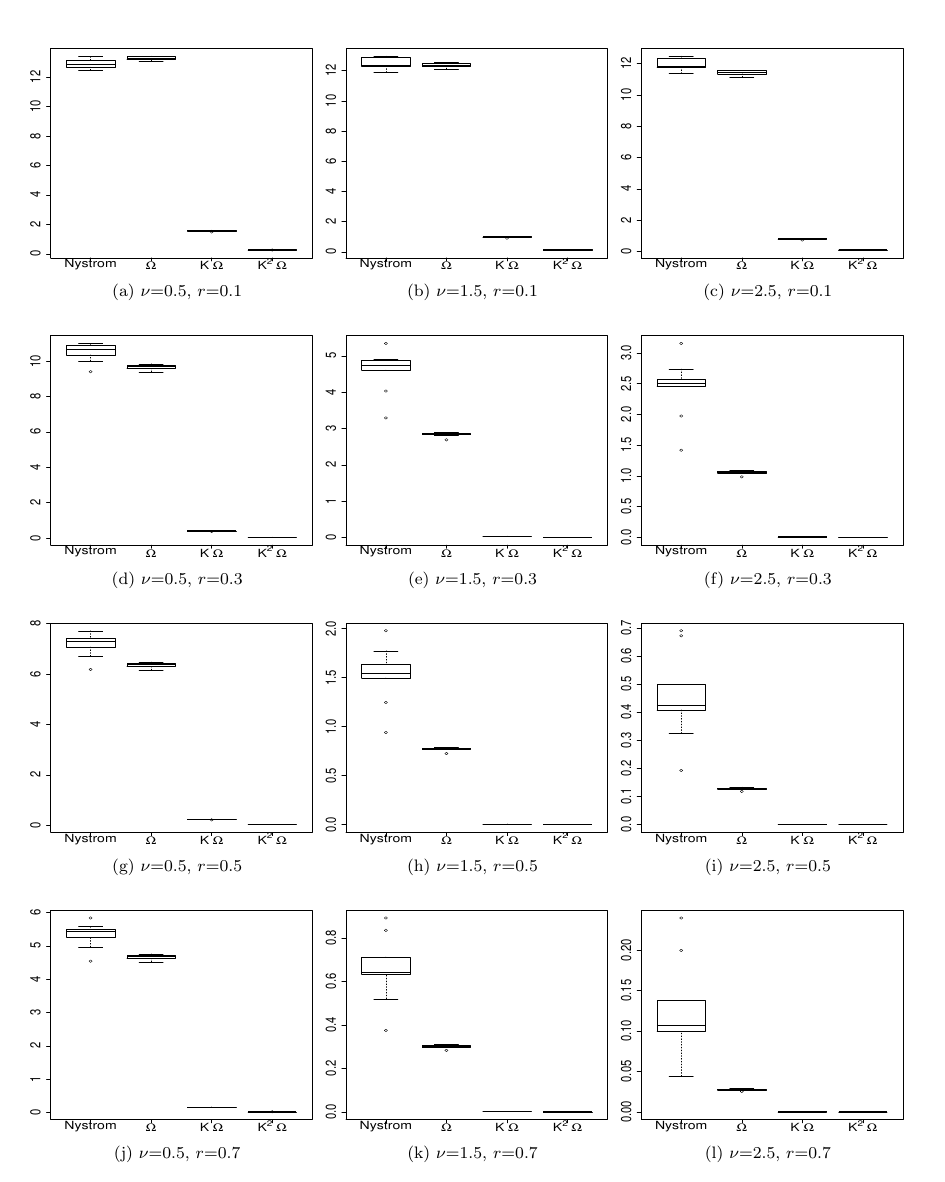}
		\caption{$ L^2 $ distance between the first 100 approximated and true eigenvalues under different smoothness $\nu$ and effective range r. Introducing randomness in $\Phi=K^\alpha\Omega$ improves the Nyst\"{o}m approximation. Letting $ \Phi $ to be $ K^\alpha\Omega $ with small power $ \alpha =1,2 $ further improves approximation}
		\label{fig:eigenvdist}
	\end{figure}

	Here we compare eigencomponent approximation performance for increasing smoothness $ \nu =0.5, 1.5, 2.5 $ and increasing spatial dependence with effective range $ r=0.1, 0.3, 0.5, 0.7 $. Figure \ref{fig:eigendist} shows the distance between the subspaces generated by the first 100 approximated and true eigenvectors. Figure \ref{fig:eigenvdist} shows the $ L^2 $ distance between the first 100 approximated and true eigenvalues. Our conclusion here is the same as in the manuscript. Introducing random matrix $ \Omega $ improves the approximation. Taking $ \Phi $ to be $ K^\alpha\Omega $ further improves approximation, where in practice $ \alpha=1 $ appears to be a good choice.
	
	\subsection*{S.2 Full Conditionals for Projection-Based Approaches}
	
	The joint posterior distribution for the full model with random projection is $ \pi(\bs{\delta},\bs{\beta},\sigma^2,\phi\mid\bs{Z} ) \propto f\left( \bs{Z} \mid\bs{\delta},\bs{\beta},\sigma^2,\phi\right) \times f(\bs{\delta}\mid\sigma^2,\phi)\times \pi(\bs{\beta}) \times \pi(\sigma^2)\times \pi(\phi)$. From this we derive the full conditionals, shown below, which can be easily sampled using one-variable-at-a-time Metropolis-Hasting algorithm. 
	\begin{equation*}
	\begin{aligned}
	\bs{\beta}\mid\bs{\beta}_- &\propto \prod_{i=1}^n f\left( Z_i\mid\bs{\beta},U_m,D_m,\bs{\delta}\right)\times\pi(\bs{\beta}),\\
	\sigma^2\mid\sigma^2_- &\propto \left( \sigma^2\right) ^{-m/2} \exp\left(-\frac{1}{2\sigma^2}\bs{\delta}^T\bs{\delta}\right)\times\pi(\sigma^2),\\
	\phi\mid\phi_- &\propto \prod_{i=1}^n f\left( Z_i\mid\bs{\beta},U_m,D_m,\bs{\delta}\right)\times \exp \left(  \frac{1}{2\sigma^2} \bs{\delta}^T\bs{\delta}\right) \times\pi(\phi),\\
	\bs{\delta}\mid\bs{\delta}_- &\propto \prod_{i=1}^n f\left( Z_i\mid\bs{\beta},U_m,D_m,\bs{\delta}\right)\times\exp \left( - \frac{1}{2\sigma^2} \bs{\delta}^T\bs{\delta}\right).
	\end{aligned}
	\end{equation*}
	The full conditionals for the restricted model with random projection is similar to the above except that $ U_m $ is replaced by $ P^{\perp}_{[X]}U_m $. 
	\subsection*{S.3 Simulation Study Results}
	
\begin{figure}[hpt]
		\centering
		\includegraphics[scale= 0.5]{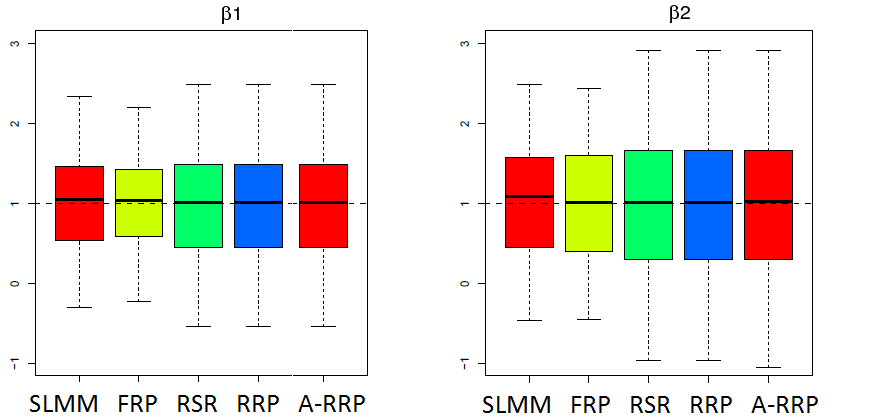}             
		\caption{Distribution of posterior mean estimates of $ \bs{\beta} $ among four models and with adjustments. The distributions all center around the true value and are comparable. Random projection models FRP and RRP with rank=50 produce results that are similar to the models they approximate.}
		\label{fig:InferenceBetaLm400}
	\end{figure}

\begin{figure}[hpt]
		\centering
		\includegraphics[scale = 0.5]{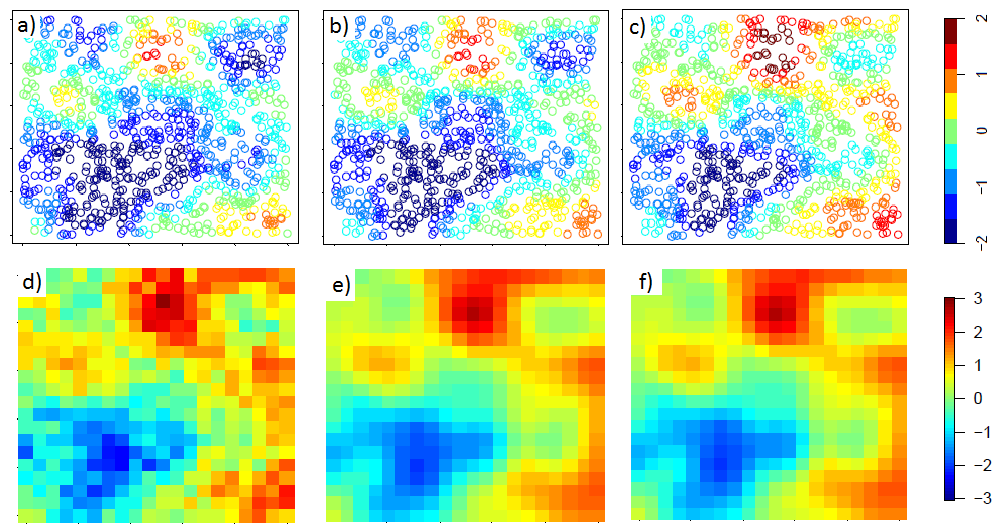}             
		\caption{Linear case with $ n = 1000 $. First row shows the random effects estimate at training locations. Second row shows the prediction on a $ 20\times 20 $ grid using random projection models. Left column is simulated data, middle column shows the results from FRP, and right column shows the results from RRP. Random projections approach works well in recovering the true random effects.}
		\label{fig:lm_pred1000}
	\end{figure}

	For the linear case, we simulate 100 data sets from the spatial linear mixed model (confounded simulation scheme) for data sizes of $ n=400 $ and $ n=1000 $. For the smaller data size, we fit both of our projection-based approaches, the spatial linear mixed model and restricted spatial regression model for overall comparisons. The distribution of $ \bs{\beta} $ estimates all center around the true value and are comparable among all four models (Figure \ref{fig:InferenceBetaLm400}); inference and prediction provided by our projection-based approaches are similar to the original models they approximate (Table \ref{tab:lm400_2.5_confound}). For the larger data size we fit both FRP and RRP with rank $ m = 50 $, which is selected based on our heuristic described in the main text. Figure \ref{fig:lm_pred1000} shows the estimated random effects at the training locations and the predicted observations at the testing locations. We see that our projection-based approaches work well in recovering the spatial patterns.
	
	\begin{table}[hpt]
		\caption{Model comparisons for linear case with $ n = 400 $.}
		\label{tab:lm400_2.5_confound}
		\centering
		\begin{tabular}{|c||l|l||l|l|l|}
			\hline
			& SLMM & FRP & RSR & RRP & A-RRP\\
			\hline
			$ \beta_1 $ (coverage) & 1.01 (0.99) & 0.98 (0.97) & 1.00 (0.07) & 1.00 (0.07) & 1.00 (0.97) \\ 
			$ \beta_1 $ mse & 0.39 & 0.46 & 0.79 & 0.79 & 0.79\\ 
			$ \beta_2 $ (coverage) & 1.02 (0.95) & 1.01 (0.95) & 1.02 (0.03) & 1.02 (0.03) & 1.02 (0.94) \\ 
			$ \beta_2 $ mse & 0.60 & 0.59 & 1.06 & 1.06 & 1.06\\ 
			$ \phi $ & 0.21 & 0.22 & 0.21 & 0.21& NA\\ 
			$ \phi $ mse & 0.62 & 0.61 & 0.62 & 0.63& NA\\  
			$ \sigma^2 $  & 1.25 & 1.34 & 1.24 & 1.20 & NA\\ 
			$ \sigma^2 $ mse & 1.32 & 1.54 & 1.26 & 1.18 & NA\\ 
			pmse & 0.13& 0.13& 0.13 & 0.13& NA\\ 
			\hline
		\end{tabular}
	\end{table}
	
\begin{figure}[hpt]
		\centering
		\includegraphics[scale= 0.6]{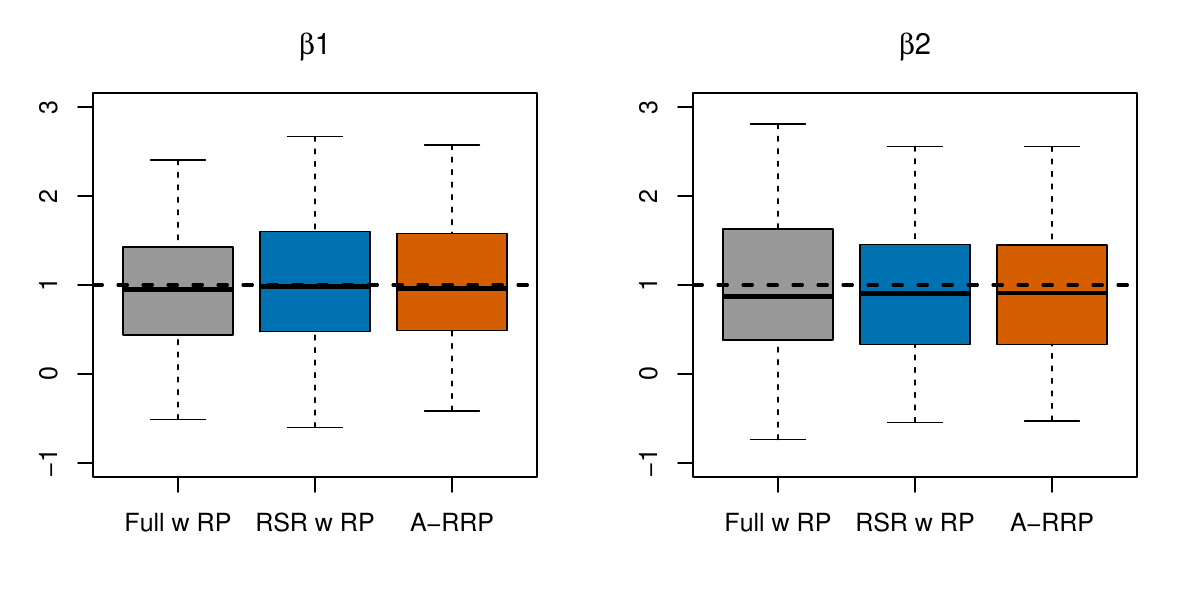} \\
		\includegraphics[scale=0.6]{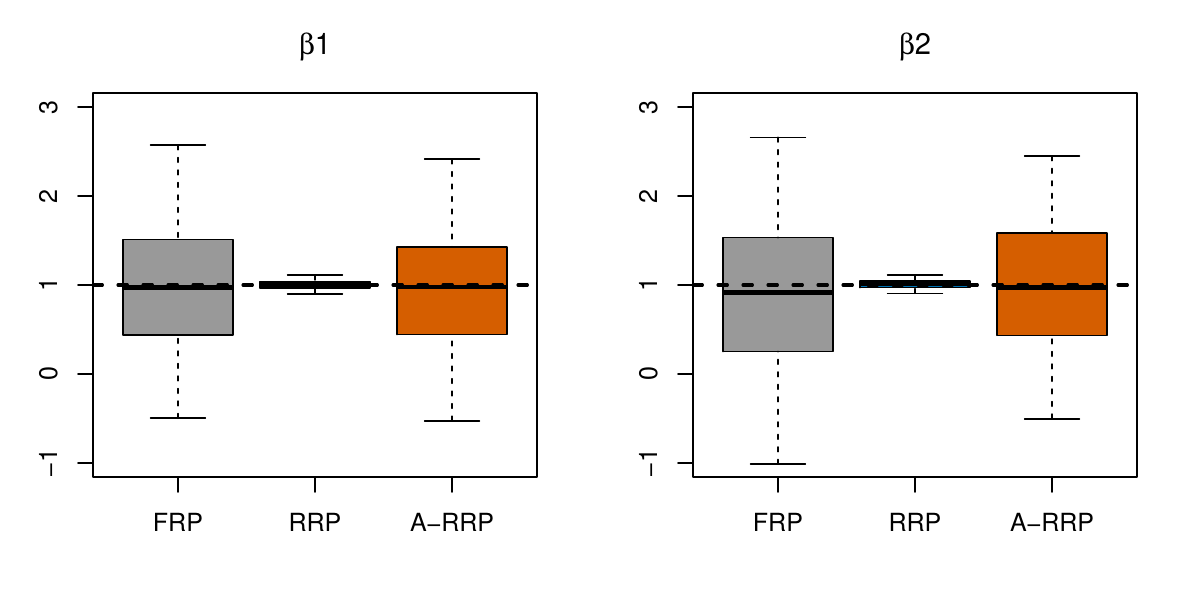}        
		\caption{Poisson simulation study: distribution of $ \bs{\beta} $ posterior mean estimates for RP models and after adjustment. First row for the confounded case, and second row for the orthogonal case. All distributions center around the true value. For the confounded case (top row), FRP and RRP have similar results; while under the orthogonal case (bottom row), RRP produce much tighter distribution. }
		\label{fig:InferenceBetaPoi1000} 
	\end{figure}

	For the Poisson case, we simulate 100 data sets from the spatial linear mixed model (confounded scheme) and restricted spatial regression model (orthogonal schemes) for data sizes of $ n=1000 $. Under the confounded simulation scheme, FRP and RRP have similar distributions for point estimates (Figure \ref{fig:InferenceBetaPoi1000}); however, the credible interval(CI) of RRP is inappropriately narrow with length 0.246(0.166, 0.367) and a coverage of 14 \% compared to the FRP, the CI of which has length 3.123(1.664, 5.743) and a coverage of 91\%. Under the orthogonal simulation scheme, RRP performs much better than FRP; its point estimates are closely centered around the true value (\ref{fig:InferenceBetaPoi1000})), its CI is 0.225(0.176, 0.299),  much narrower compared to 2.985(1.666, 4.995) of the FRP, and both RRP and FRP have coverages that are comparable to the nominal rate. Under both simulation schemes, the adjusted inference A-RRP provides similar results to FRP. Hence, we can fit only the RRP model in practice for its computational efficiency, then apply the adjustment to recover inference results for the full model.

	\subsection*{S.4 A Comparison with an Existing Method for Areal Data}
\begin{figure}[hpt]
		\centering
		\includegraphics[scale=0.28]{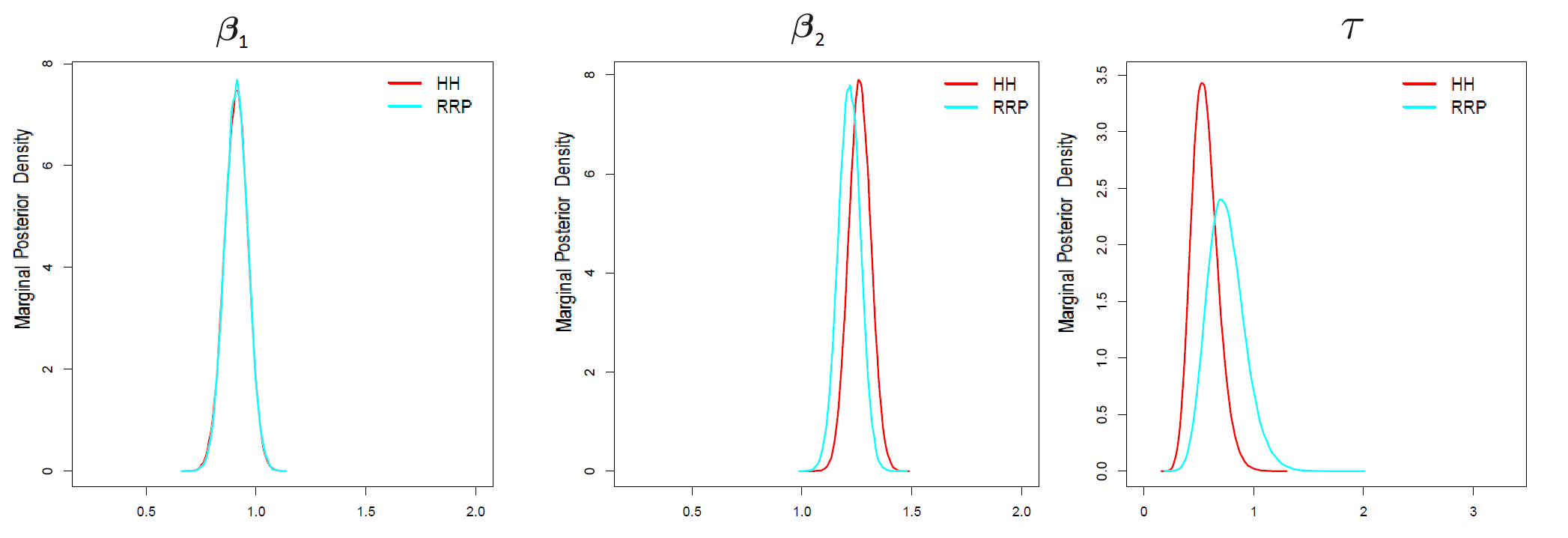}
		\caption{Marginal posterior plots for HH and RRP models. Results from the two models are comparable. RRP = restricted model with random projection}
		\label{fig:CompHH}
	\end{figure}

	Here we compare our approach with an existing method for lattice/areal data \citep{HughesHaran}. We simulate a count data set with $ n = 900,\tau=1  $ from: 
	\begin{equation}
	\label{eqnappendix:car}
	\begin{aligned}
	g\left\lbrace E( \bs{Z}\mid \bs{\beta} )\right\rbrace &=  \bs{x_1} +\bs{x_2} + \bs{W},\\
	p(\bs{W}\mid\tau) &\propto  \tau^{rank (Q)/2} \exp \left( -\frac{\tau}{2} \bs{W}^TQ\bs{W}  \right).
	\end{aligned}
	\end{equation}
	The ICAR model has improper prior, meaning its precision matrix is rank deficient; therefore, direct simulation from \eqref{eqnappendix:car} is not feasible. Hence, the spatial random effects is simulated using the eigencomponents of the precision matirx $ Q $. Let $ (\lambda_i, e_i) $ denote the eigenpairs of $ Q $, we simulate $ \delta_i \sim N(0,\lambda_i^{(-1)}) $ for $ \lambda_i \ne 0  $. Then $ W = \sum_i \delta_i  e_i $ has the desired distribution. To reduce the dimension of $ \bs{W} $ using RRP, we will first invert $ Q $ using generalized inverse, then approximate $ Q^{-1} $ using Algorithm 1 from the main text. The full conditionals of RRP for this reparameterized model can be easily derived. We then fit both RRP and HH to the simulated data set for comparison. Figure \ref{fig:CompHH} shows that the marginal posterior density plot are similar from the two models. 
	
	\subsection*{S.5 A Comparison with Predictive Process for Point-Referenced Data}
\begin{figure}[hpt]
		\centering
		\includegraphics[scale=0.7]{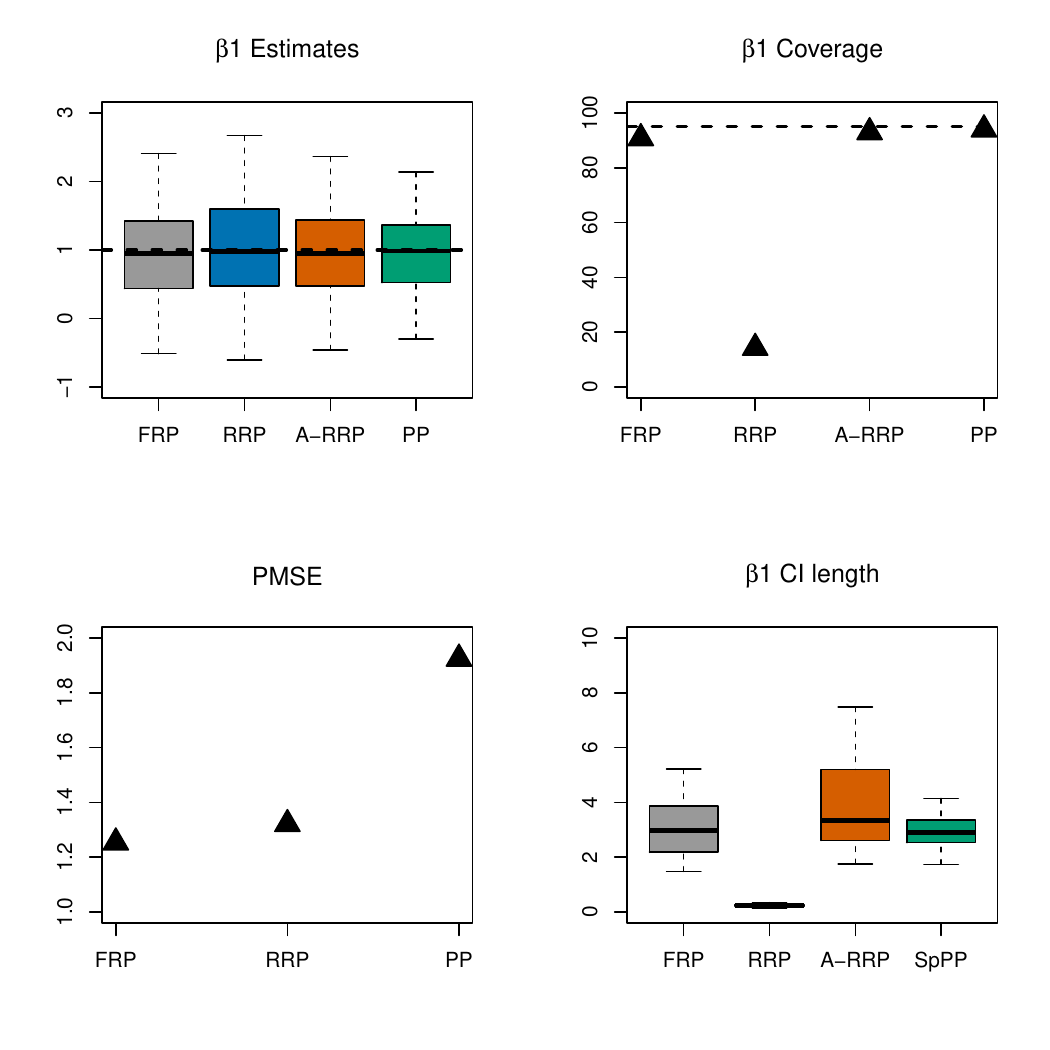}\\
		\caption{Poisson simulation study: compare projection-based models with predictive process on gridded knots. The point estimate distributions for $ \beta_1 $ are comparable (top left); while the coverage for RRP is much lower than the others, however after adjustment, the coverage for A-RRP is corrected and is comparable to FRP and PP (top right). Both FRP and RRP have better prediction performance than PP (bottom left). The length of the CIs for FRP and PP are comparable, while RRP produce much narrower CI; but after adjustment the CI gets much wider (bottom right). FRP = full model with random projection, RRP = restricted model with random projection, A-RRP = adjusted inference for RRP.}
		\label{fig:RPvsPPgrid}
	\end{figure}

	To compare the performance of our projection-based approaches with the predictive process, we simulate 100 Poisson data sets from the traditional SGLMM. We fit both FRP and RRP with rank $ m=50 $ to the datasets, and compare their results with the predictive process with reference points on a $ 7\times7 $ grid. In this simulation study, our projection-based approaches provide comparable inference and smaller mean prediction square error (MPSE) (Figure \ref{fig:RPvsPPgrid}).
	
	\subsection*{S.6 SGLMMs with small-scale (nugget) spatial effect}
	For SGLMMs where inclusion of small scale, non-spatial heterogeneity is appropriate, the model becomes, \begin{equation}\label{eqn: heterspatial}
	g\left\lbrace E(Z(s)\mid \beta,W(s))\right\rbrace = X(s) \beta + w(s) + \epsilon(s),
	\end{equation}where $ \epsilon(s)\stackrel{iid}{\sim} N(0,\tau^2) $. 
	We provide implementations of our method for two cases: (1) when Gibbs sampling of the latent variables is available, and (2) when it is not. Examples for case (1) are the spatial binary model with probit link \citep[considered by][]{Berrett2016} and spatial probit model for correlated ordinal data \citep{Schliep2015};  examples for case (2) are already considered in this manuscript. 
	
	We begin by redefining some notation. Let $ \bs{W} = (W_1,\dots, W_n)^T $ denote the latent variable, $ \bs{Z} = (Z_1,\dots,Z_n)^T $ the observed spatial binary data and $ X $ the $ n\times p $ design matrix.
	
	\textbf{Case (1):} We first consider the case where Gibbs sampling is available for the latent variables, for example when using SGLMM with a probit link for binary data. The model is defined as
	\begin{equation}
	Z_i = \begin{cases}
	1, & Y_i \ge 0 \\
	0, & Y_i < 0
	\end{cases}
	\end{equation}
	where $ Y_i = X_i\beta + W_i + \epsilon_i$. $ \bs{W} \sim  MVN(0, \sigma^2R_\phi)$ captures large-scale spatial variation and $ \epsilon_i \stackrel{i.i.d}{\sim} N(0,\tau^2)  $ captures small-scale variation. The conditional distribution for $ Y\mid\bs{\beta},\sigma^2,\phi,\tau^2 $ is therefore multivariate normal with mean $ X\bs{\beta}$ and variance $ \sigma^2R_\phi + \tau^2I $. \textbf{Our method can be used to facilitate model fitting in this case as follows:} We approximate the eigen-components of $ R_\phi $ using random projections and obtain its first $ m $ eigenvectors $ U_\phi = [\bs{u}_1,\dots,\bs{u}_m]$ and eigenvalues $D_\phi = \text{diag}(\lambda_1, \dots, \lambda_m)$. Let $ M_\phi = U_\phi D_\phi ^{1/2} $ be the projection matrix, then we reduce the dimension of the latent variables by approximating $ \bs{W}  $ with $ M_\phi \bs{\delta}$. For a specific value of $ \phi $, we can treat $ M_\phi $ as fixed spatial covariates and $ \bs{\delta} $ the corresponding coefficients. Write $ X_\phi =[X, M_\phi] $ and $ \bs{\beta}_\phi = (\bs{\beta}^T, \bs{\delta}^T)^T$ as the reparameterized design matrix and coefficients, respectively, then $ {Y_i}$ is approximated by $ X_i\bs{\beta} + M_{\phi,i} \bs{\delta} +\epsilon_i $ and can be rewritten as $ X_{\phi,i}\bs{\beta}_\phi + \epsilon_i$. We use a normal conjugate prior for $ \bs{\beta} $, inverse gamma conjugate priors for $ \sigma^2 $ and $ \tau^2 $, and a uniform prior for $ \phi $. Then, fitting the reduced-rank Bayesian probit model involves the following steps.
	\begin{itemize}
		\item[]At the $ t^{th} $ iteration of the algorithm,
		\item[] Step 1: Gibbs update for latent variables. Sample $ \bs{{Y}}^{(t)} $ from $ \bs{{Y}}|\bs{Z},\bs{\beta}^{(t-1)},{\sigma^2}^{(t-1)},\phi^{(t-1)},{\tau^2}^{(t-1)}$
		\begin{itemize}
			\item[(a)] Compute projection matrix $ M_\phi $ for $ \phi^{(t-1)} $. Form $ X_\phi $ and $ \bs{\beta}_\phi $.
			\item[(b)] For $ i = 1,\dots, n $, draw $ Y_i $ from
			\begin{equation*} {Y_i}|\bs{Z}, \bs{Y_{-i}},\bs{\beta},\sigma^2,\phi,\tau^2 \sim \begin{cases}
			TN (X_{\phi,i}\bs{\beta}_\phi,\tau^2, 0,\infty ), & \text{if } Z_i = 1 \\
			TN (X_{\phi,i}\bs{\beta}_\phi,\tau^2,-\infty,0 ), & \text{if } Z_i = 0, \\
			\end{cases}
			\end{equation*}
			where $ TN (\mu_{Y_i},\sigma^2_{Y_i}, 0,\infty ) $ is a truncated normal distribution with lower bound 0, upper bound  $ \infty $, mean $ X_{\phi,i}\bs{\beta}_\phi $ and variance $ \tau^2 $.
		\end{itemize}
		\item[] Step 2: Gibbs update for $ \bs{\beta}_\phi $.
		\begin{itemize}
			\item[]Sample from $ \bs{\beta}_\phi\mid \bs{Z}, \bs{Y}^{(t)},{\sigma^2}^{(t-1)}, \phi^{(t-1)},{\tau^2}^{(t-1)}\sim \texttt{MVN}\left(\hat{\bs{\beta}}_\phi, (\frac{1}{{\tau^2}^{(t-1)}} X_\phi^TX_\phi + \Sigma_\beta^{-1})^{-1}\right) $,\\
			\item[]where $ \hat{\bs{\beta}}_\phi = (\frac{1}{{\tau^2}^{(t-1)}} X_\phi^TX_\phi + \Sigma_\beta^{-1})^{-1} \frac{1}{{\tau^2}^{(t-1)}}X_\phi^T\bs{Y}^{(t)} $, and $ \Sigma_\beta = \left[\begin{matrix}
			\Sigma_0 & 0 \\
			0 & {\sigma^2}^{(t-1)} I_{m\times m}
			\end{matrix}\right]$ with $ \Sigma_0 $ denotes the normal prior variance.
		\end{itemize}
		\item[] Step 3: Gibbs update for $ \tau^2 $.
		\item[] Step 4: Gibbs update for $ \sigma^2 $.
		\item[] Step 5: Metropolis-Hastings update for $ \phi $.
	\end{itemize} We have not provided details for steps 3-5 since they remain the same as when fitting SGLMMs in general. Furthermore, techniques for dealing with non-identifiable parameters \citep{Berrett2012,Berrett2016} can also be used.
	
	\textbf{Case (2):} We now consider the case where Gibbs sampling from the latent variable is not available. We first explain why the reparameterization for Case (1) is not suitable here, and then provide an alternative strategy. In Case (1) above, $ \bs{W} $ is reparameterized with a low-rank representation, however, the dimension of latent variable $ \bs{Y} $ remains high; $ \bs{Y} $ is approximated by $ X\bs{\beta}+ M_\phi\bs{\delta} + \bs{\epsilon}$, and has a normal distribution with mean $ X\bs{\beta} $ and covariance $ \sigma^2M_\phi M_\phi^T + \tau^2 I$. Constructing efficient MCMC to sample $ \bs{Y} $ from its full conditional distribution is not easy due to its high dimensions. Hence, we propose an alternative: reduce the dimension of $ \bs{Y} $ by approximating $ \bs{W} + \bs{\epsilon} $ with $ U_\theta D_\theta^{1/2}\bs{\delta} $, where $ U_\theta $ and $ D_\theta$ are eigenvectors and eigenvalues of $ \sigma^2R_\phi + \tau^2 I $, respectively. Hence, the eigencomponents here depend on all parameters $ \bs{\theta} = (\sigma^2,\phi,\tau^2)^T $ of the covariance function. In fact $ U_\theta $ is identical to $ U_\phi $ from Case (1), and $ D_\theta $ is identical to $ \sigma^2D_\phi + \tau^2I_{m\times m} $. This alternative reparameterization provides some computational gains. The latent variable $ \bs{Y} $ is now approximated by $ X\bs{\beta} + M_\theta\bs{\delta}  = [X, M_\theta](\bs{\beta}^T, \bs{\delta}^T)^T $ whose full conditional distribution has $ m + p $ dimensions. Reducing the dimension of the posterior distribution allows for easier construction of efficient MCMC.